\documentclass[10pt]{article}

\usepackage[caption=false]{subfig}
\usepackage{fullpage, authblk, amsmath, amssymb, epsfig, txfonts, cite}

\usepackage{epstopdf}
\pdfoutput=1

\begin{document}
    \title{A Modular Approach to Adaptive Structures}

    \author[1]{Markus Pagitz\thanks{markus@pagitz.de}}
    \author[2]{Manuel Pagitz}
    \author[1]{Christian H\"uhne}
    \affil[1]{Institute of Composite Structures and Adaptive Systems, German Aerospace Center}
    \affil[2]{Institute of General Practice, Goethe University Frankfurt}

    \renewcommand\Authands{ and }

    \date{}

    \maketitle

    \begin{abstract}
        A remarkable property of nastic, shape changing plants is their complete fusion between actuators and structure. This is achieved by combining a large number of cells whose geometry, internal pressures and material properties are optimized for a given set of target shapes and stiffness requirements. An advantage of such a fusion is that cell walls are prestressed by cell pressures which increases, decreases the overall structural stiffness, weight. Inspired by the nastic movement of plants, Pagitz et al. 2012 Bioinspir. Biomim. 7 published a novel concept for pressure actuated cellular structures. This article extends previous work by introducing a modular approach to adaptive structures. An algorithm that breaks down any continuous target shapes into a small number of standardized modules is presented. Furthermore it is shown how cytoskeletons within each cell enhance the properties of adaptive modules. An adaptive passenger seat and an aircrafts leading, trailing edge is used to demonstrate the potential of a modular approach.

        \vspace{3mm}\textbf{Keywords}\ \ \ \ \textit{adaptive - biomimetic  - cellular - compliant - cytoskeleton - modular - morphing - structures}
    \end{abstract}
    \vspace{10mm}


    \noindent \large\textbf{Nomenclature}

    \footnotesize
    \begin{align*}
        \begin{array}{p{1.5cm}l}
            $\alpha$, $\alpha^-$, $\alpha^+$    & \textrm{central, lower, upper limit angle}\\
            $\theta$                            & \textrm{sum of central angles}\\
            $\xi$                               & \textrm{dimensionless coordinate}\\
            $\rho$, $\sigma_y$                  & \textrm{material density, yield strength}\\
            $\Pi$                               & \textrm{error norm}\\\\
            $\mathbf{f}$, $\mathbf{K}$          & \textrm{first, second derivative of error norm}\\
            $h$, $r$, $s$                       & \textrm{height, radius, chord length of circular segment}\\
            $i$, $j$                            & \textrm{integers}\\
            $m$, $n$                            & \textrm{\#modules, \#different modules}\\
            $p$                                 & \textrm{pressure}\\
            $q$                                 & \textrm{refinement level}\\
            $t$                                 & \textrm{thickness}\\
            $\mathbf{v}$                        & \textrm{vector}\\
            $w$                                 & \textrm{weight}\\
            $x$, $y$                            & \textrm{cartesian coordinates}\\\\
            $A$, $\mathbf{A}$                   & \textrm{cross sectional area, adaptive module}\\
            $\mathbf{C}$                        & \textrm{center of circular arc}\\
            $F$                                 & \textrm{axial force}\\
            $L$                                 & \textrm{arc length of adaptive module, cell side length}\\
            $M$, $\mathbf{M}$                   & \textrm{moment, mechanical module}\\
            $\mathbf{P}$, $\mathbf{Q}$          & \textrm{end, central node of module}\\
            $\mathbf{R}$                        & \textrm{rigid module}\\
            $\mathbf{S}$                        & \textrm{structure made from various modules}
        \end{array}
    \end{align*}
    \newpage
    \normalsize


    \section{Introduction}
\begin{figure}[htbp]
            \begin{center}
                \includegraphics[width=0.9\textwidth]{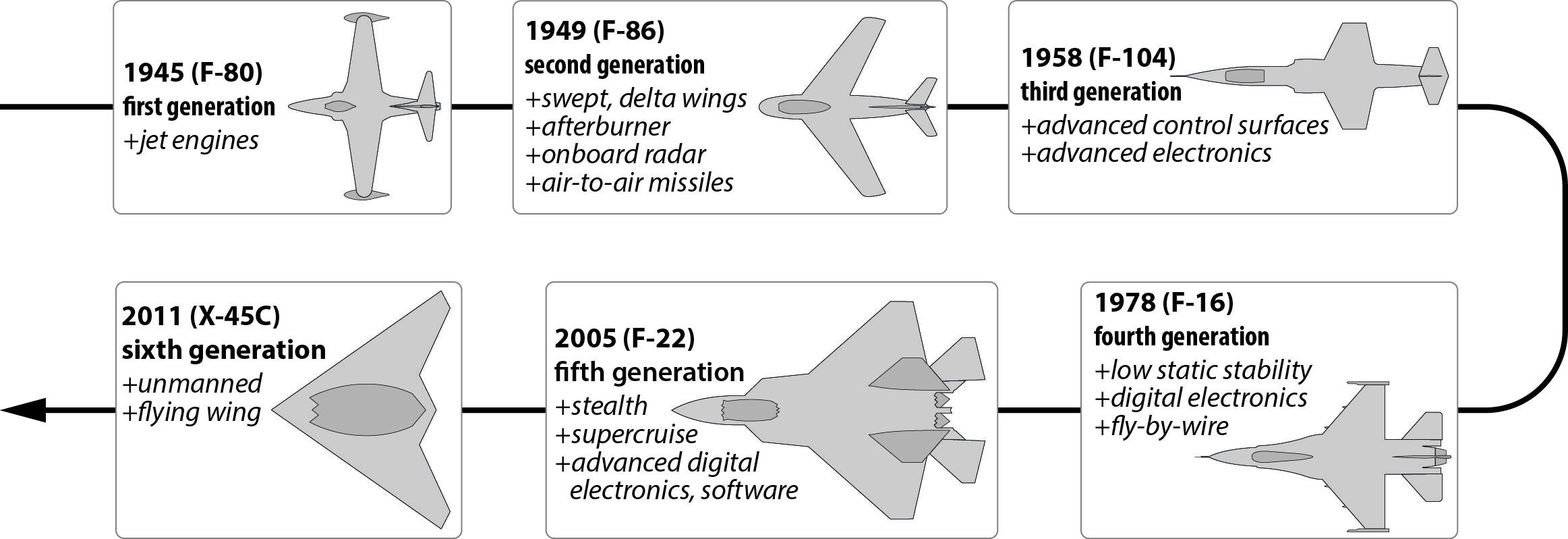}
                \caption{Fighter jet development since Second World War.}
                \label{pic:Figure_1_1}
            \end{center}
        \end{figure}

        Advances in electronics, computer aided engineering, manufacturing and material technology had a great impact on machine design. A good example is the development of fighter jets that spans seven decades and started with propeller machines that were upgraded with simple jet engines. In contrast, forthcoming fighter jets are unmanned flying wings with stealth properties and advanced electronics, Figure~\ref{pic:Figure_1_1}. Improvements in the aerodynamic design were mostly driven by the rapid development of computers that allowed the use of unstable designs. The relaxation of aerodynamic design constraints is continued by the current shift towards unmanned aircraft. In contrast, changes in the structural design were not as radical. For example, control surfaces are still rigid bodies that require gaps in an aircrafts skin and thus create a source of radar reflections. This is a limiting factor as aircraft become increasingly stealthy. A remedy would be the use of adaptive structures that eliminate these gaps and additionally expand the optimal flight regime. An overview of potential technologies for adaptive structures can be found in \cite{Vasista2012}. Note that the applicability of adaptive structures is by no means limited to aircraft. Other application areas are, for example, architectural elements such as seats and couches.

\begin{figure}
            \begin{center}
                \begin{minipage}[c]{0.35\textwidth}
                    \subfloat[]{
                        \includegraphics[width=1\textwidth]{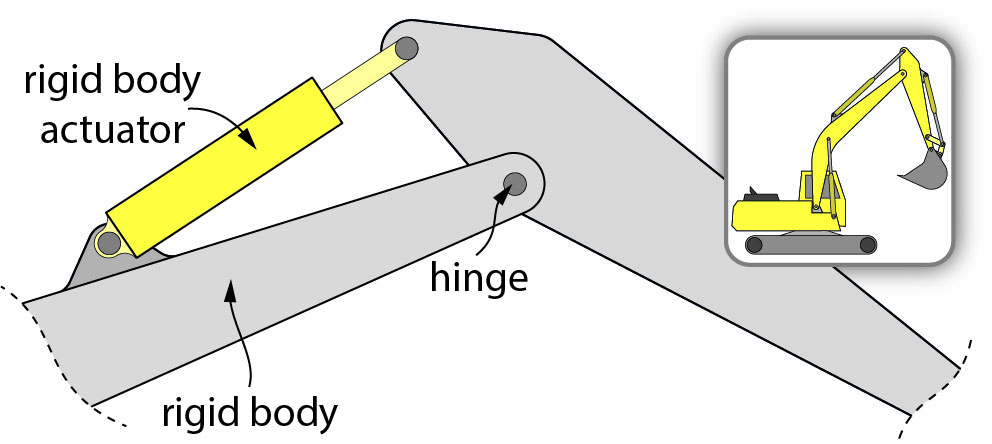}}

                    \subfloat[]{
                        \includegraphics[width=1\textwidth]{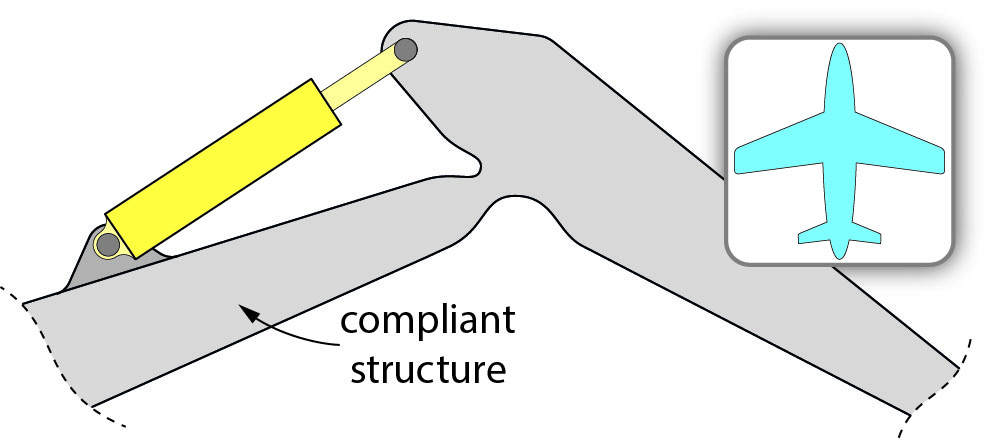}}

                    \subfloat[]{
                        \includegraphics[width=1\textwidth]{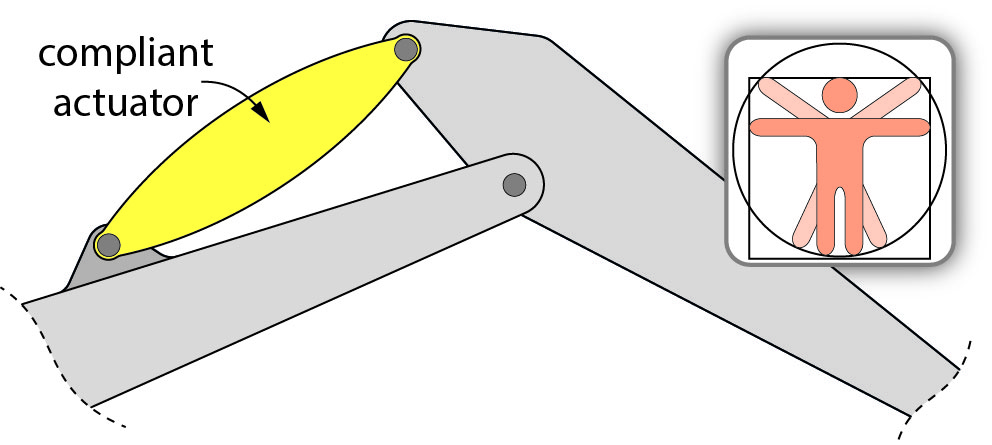}}

                    \subfloat[]{
                        \includegraphics[width=1\textwidth]{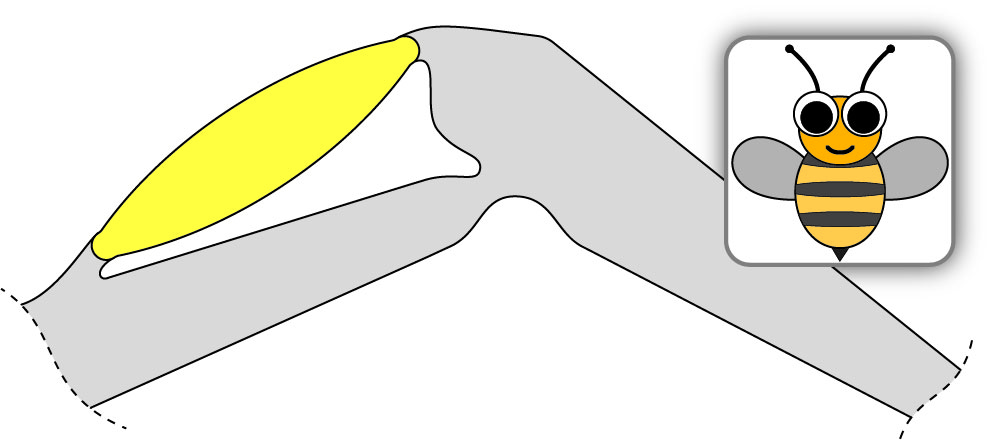}}
                \end{minipage}\hspace{20mm}
                \begin{minipage}[c]{0.35\textwidth}
                    \subfloat[]{
                        \includegraphics[width=1\textwidth]{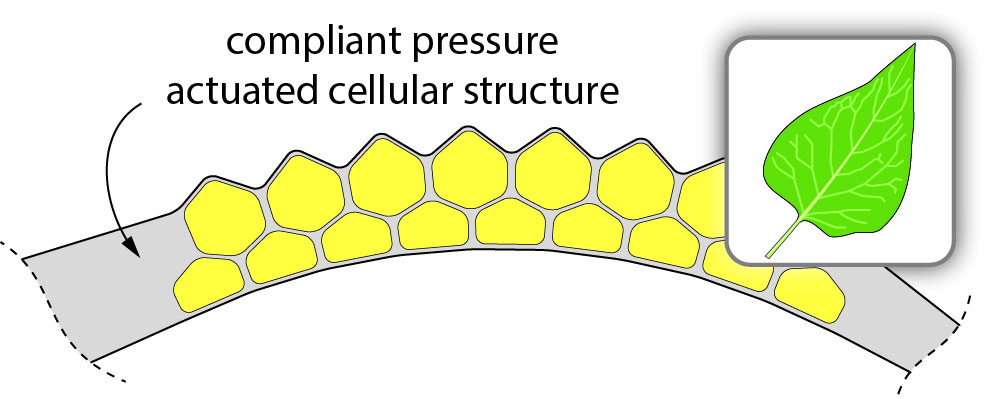}}\vspace{2mm}
                    \caption{\footnotesize
                            A mechanism with a single rotational degree of freedom can be realized by using\ldots \vspace{3mm}\newline
                            (a) \ldots a rigid body actuator and structure. This is the classical design principle for machines such as excavators and aircraft~\cite{Kintscher2012}. \vspace{3mm}\newline
                            (b) \ldots a rigid body actuator and a compliant structure. Compliant structures reduce the number of parts and potentially reduce the total weight and manufacturing cost. This design principle is currently considered for adaptive aircraft wings~\cite{Kota2009}. \vspace{3mm}\newline
                            (c) \ldots a compliant actuator and a rigid body structure. The locomotor system of humans and many animals is based on this design principle. \vspace{3mm}\newline
                            (d) \ldots a compliant actuator and structure. This design principle is used, for example, for the wings of bees~\cite{Vigoreaux2006}. The lack of hinges reduces backlash, friction and wear. \vspace{3mm}\newline
                            (e) \ldots a design principle that does not distinguish between actuators and structure. A compliant pressure driven actuator is formed by a cellular structure. Cell pressures introduce a prestress into the structure which increases its stiffness~\cite{Pagitz2012,Pagitz2013,Pagitz2014-1,Pagitz2014-2}. Nastic plants are based on this design principle.}
                    \label{pic:Figure_1_2}
                \end{minipage}
            \end{center}
        \end{figure}

        Kinematic mechanisms are the backbone of adaptive structures. Different design principles for a directly actuated, single degree of freedom mechanism are shown in Figure~\ref{pic:Figure_1_2}. Both, a compliant and rigid body actuator, structure are considered\footnote{For example, hydraulic cylinders, electric motors are rigid body actuators whereas pneumatic muscles, electroactive polymers are compliant actuators.}. Advantages of compliant actuators, structures are the reduced number of parts as well as the lack of backlash, friction and wear. On the other hand, these advantages come at the cost of an increased design complexity since elastic stresses and fatigue have to be considered. Hence, possible rotation angels are limited. Most actuators possess only a single degree of freedom. This often causes a kinematic incompatibility between mechanism and actuators which prevents a complete fusion between actuators and structure. Nastic plants are an exception insofar that there exists no separation between actuators and structure~\cite{Pagitz2012,Pagitz2013}. They are made from a large number of cells whose geometry, internal pressures and material properties are tailored for a given set of target shapes and stiffness requirements~\cite{Pagitz2014-1}. This leads to a synergistic effect since cell pressures introduce a prestress into the structure which increases, decreases the overall stiffness, weight. Compliant pressure actuated cellular structures with fixed properties can be made from single materials that range from elastomers to metals~\cite{Pagitz2014-2}. This allows the use of advanced manufacturing techniques such as injection molding or rapid prototyping. The choice of cell materials affects cell sizes and thus the required cell pressures. Hence it is possible to design cellular structures for a wide pressure range.\newline

        The need for \textit{advanced aircraft designs} is one of the driving forces behind research on plant inspired adaptive structures. Although not directly related to adaptive structures, Khire et al.~\cite{Khire2006} published an interesting article about inflatable structures that consist of a large number of uniformly pressurized hexagonal cells with a regular, prismatic geometry. Vos et al.~\cite{Vos2011} subsequently published and patented a similar concept. They combine a large number of regular, prismatic cells with an elastic plate such that its deformation state depends on cell pressures. Inspired by the fibrillar network in plant cell walls, Philen et al.~\cite{Philen2007} proposed a concept for smart materials that are based on a large number of pressurized artificial muscles~\cite{Zhang2012}. Vasista and Tong~\cite{Vasista2013} used topology optimization to compute cell geometries for desired deformations of an adaptive structure. A summary of recent work in this field that was supported by the Defense Advanced Research Agency, National Science Foundation and the United States Army can be found in the book edited by Wereley and Sater~\cite{Wereley2012}.\newline

        Another driving force is the emerging field of \textit{soft robotics}. Most actuators for soft robots are made from a single row of regular cells that are not optimized for any target shape~\cite{Deimel2013,Marchese2014,Shepherd2011}. Hence, materials that can undergo large elastic strains are used to compensate for the simple cell geometries. These kind of structures can be considered to be a special case of~\cite{Pagitz2012}. A modular approach to soft robotics that is based on regular, single row actuators was published by Onal and Rus~\cite{Onal2012}. The aim of this article is to introduce pressure actuated cellular modules that consist of two optimized cell rows and connectors at both ends (Section~2). Furthermore, it is shown how cytoskeletons can be used to decrease, increase the weight, stiffness of adaptive modules. An algorithm that breaks down any continuous target shapes into a set of modules is presented (Section~3). It is shown that a small number of adaptive, mechanical and rigid modules can be used to construct adaptive structures such as airfoils and passenger seats (Section~4).


    \section{Modules}
        \subsection{Adaptive Modules}
\begin{figure}[htbp]
                \begin{center}
                    \subfloat[]{
                        \includegraphics[height=0.34\textwidth]{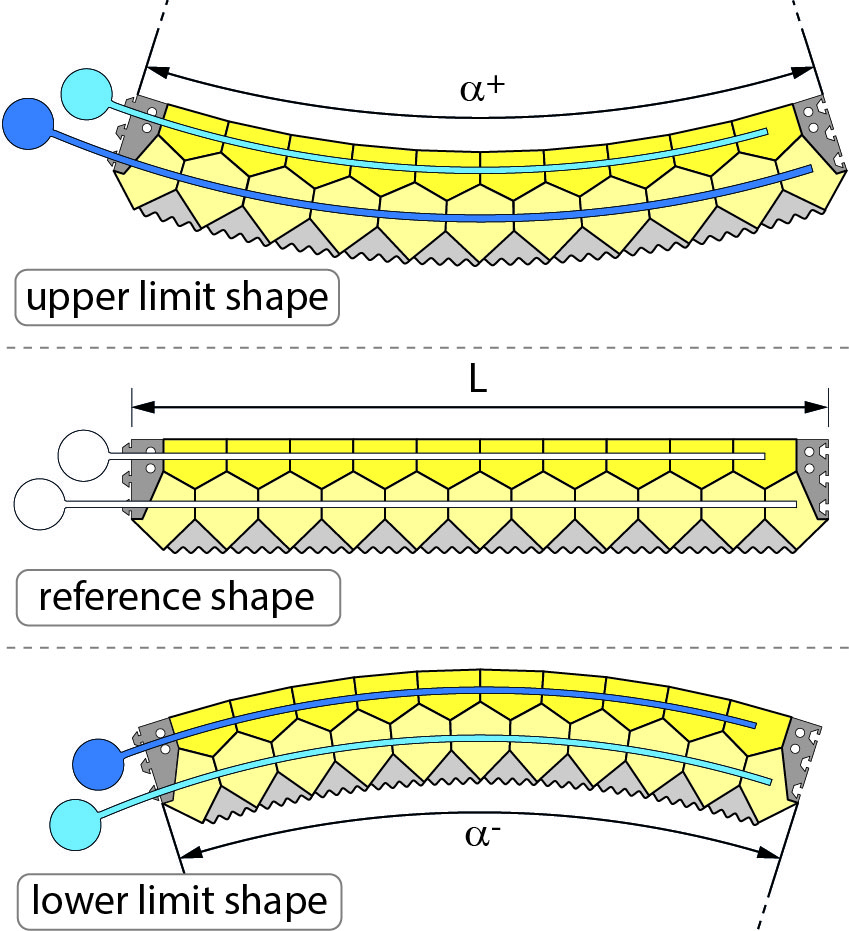}}\hspace{5mm}
                    \subfloat[]{
                        \includegraphics[height=0.34\textwidth]{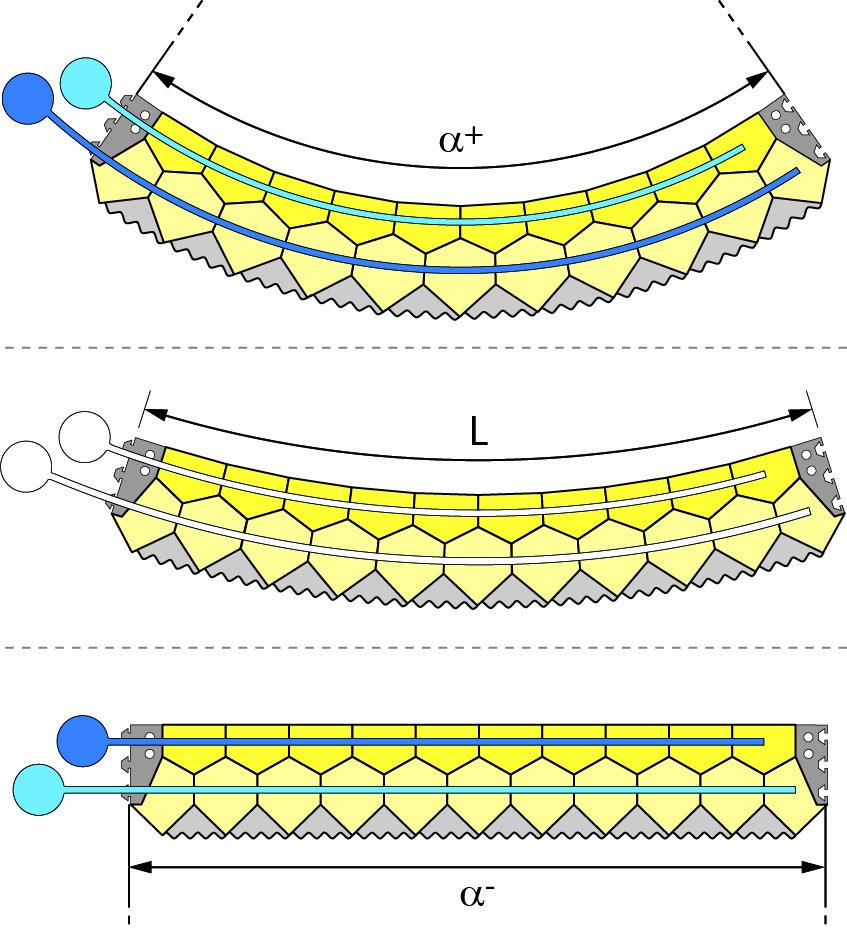}}\hspace{5mm}
                    \subfloat{
                        \includegraphics[height=0.34\textwidth]{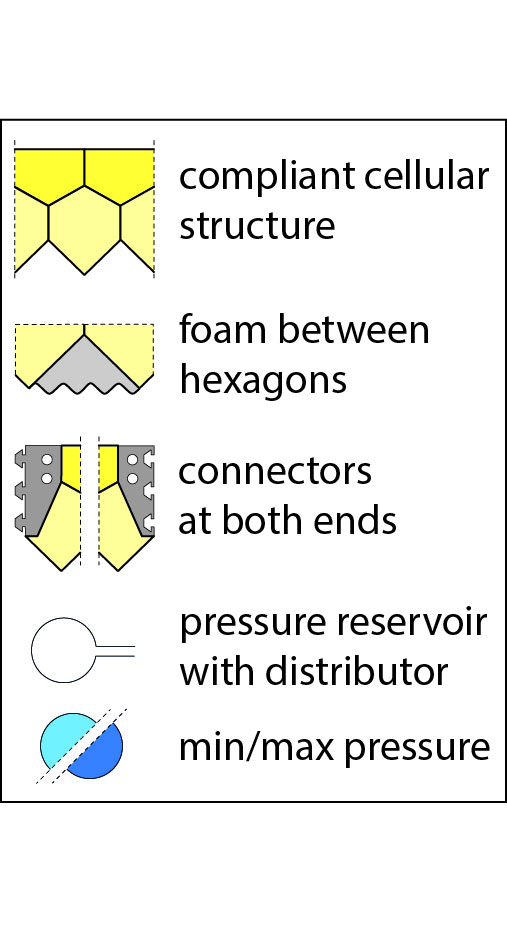}}
                    \caption{An adaptive module $\mathbf{A}_{\alpha^-,\alpha^+}^L$ with an arc length $L$ can continuously change its shape between central angles $\alpha^-$ and $\alpha^+$ by varying the pressure ratio between cell rows. Reference and limit shapes of an adaptive module for (a) $\mathbf{A}^L_{\mp 35^\circ}$ and (b) $\mathbf{A}^L_{0^\circ,70^\circ}$.}
                    \label{pic:Figure_2_1}
                \end{center}
            \end{figure}

            Pressure actuated cellular structures are usually purpose made for a given set of target shapes, cell pressures and stiffness requirements. Hence it is necessary to go through the whole design and manufacturing process for every prototype. Such a procedure is time consuming and expensive. Costs during the initial prototyping phase can be reduced by using a modular approach. A sketch of two different adaptive modules is shown in Figure~\ref{pic:Figure_2_1}. It can be seen that a module consists of two cell rows and connectors at both ends that enable an arbitrary combination of modules. Furthermore, the space between adjacent hexagons is filled with a flexible foam to obtain a smooth inner surface.

            It is subsequently assumed that equilibrium shapes of adaptive modules are circular arcs\footnote{The cellular structure of an adaptive module can be designed for any set of continuous target shapes.}. This choice is motivated by the fact that, apart of edge effects, regular cellular structures deform into circular arcs. Hence, cell side lengths and rotation angles at cell corners are more uniform across a module. The used notation for a single adaptive module is $\mathbf{A}_{\alpha^-, \alpha^+}^L$ where $L$ is the arc length and $\alpha^-$, $\alpha^+$ are the central angles of the lower, upper limit shape. Note that an adaptive module can continuously change its shape between both limit shapes by altering the pressure ratio between cell rows. Adaptive modules with a given arc length can be designed for arbitrary limit angles. However, the shape changing capability $\Delta \alpha=\alpha^+-\alpha^-$ of a module is proportional to the required minimum number of cells. Therefore, the module stiffness is inversely proportional to the shape changing capability since the module thickness is inversely proportional to the number of cells.


        \subsection{Cytoskeletons}
            The proposed concept for pressure actuated cellular structures mimics the nastic movement of plants by combining rows of prismatic cells with tailored cell side lengths and thicknesses. The overall weight and stiffness of these structures can be further improved by using cytoskeletons. A similar approach can be found in nature. For example, the shape of eukaryotic cells is determined by cytoskeletons. Ingber~\cite{Ingber1993} showed that the complex structure of cytoskeletons might be explained on the basis of tensegrity structures~\cite{Pagitz2009}. His main argument is that local remodeling phenomena of cytoskeletons must be seen in a global, cellular context. In order to simplify matters, two different kind of cytoskeletons are subsequently considered.


            \subsubsection{Stiffness}
\begin{figure}[htbp]
                    \begin{center}
                        \subfloat[]{
                            \includegraphics[height=0.37\textwidth]{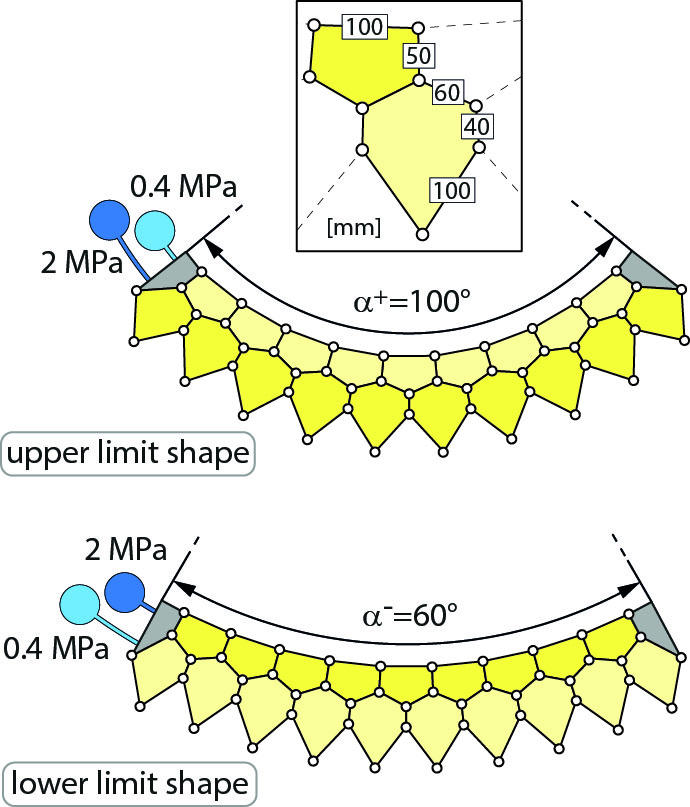}}\hspace{2mm}
                        \subfloat[]{
                            \includegraphics[height=0.37\textwidth]{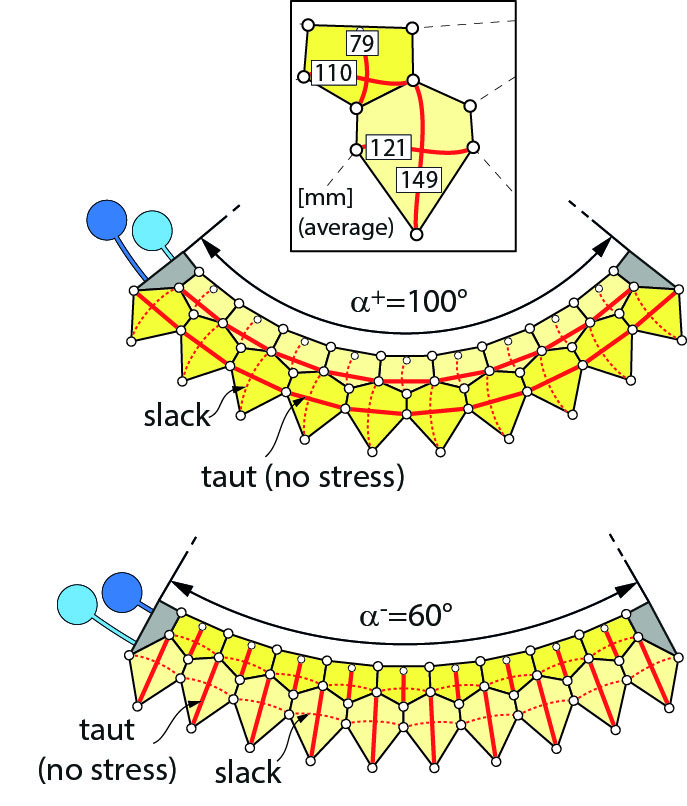}}\hspace{2mm}
                        \subfloat[]{
                            \includegraphics[height=0.37\textwidth]{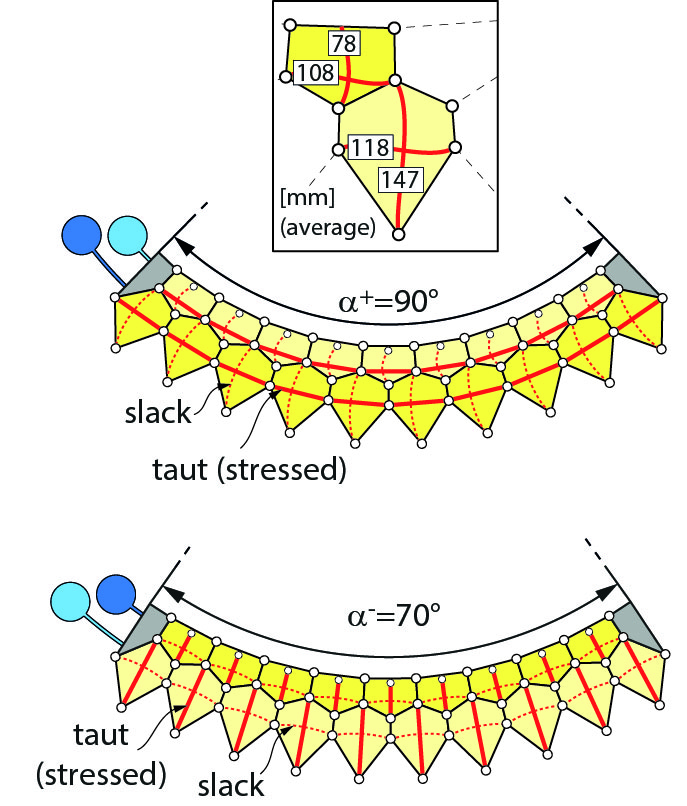}}
                        \caption{Adaptive module (a) without and (b-c) with cytoskeletons that are (b) unstressed, (c) stressed at limit shapes.}
                        \label{pic:Figure_2_2}
                    \end{center}
                \end{figure}

                The first kind of cytoskeletons limits the deformability of cells. This is achieved with an internal structure that can carry large tensile and only minor compressive forces. Hence, these cytoskeletons are, depending on the deformation states of cells, either slack or taut. Three different versions of an adaptive module with identical cell geometries\footnote{Identical cell geometries are used for the loaded cantilever in Pagitz et al.~\cite{Pagitz2012}.} are considered as illustrated in Figure~\ref{pic:Figure_2_2}.

                \noindent The first version lacks any cytoskeletons and possesses a shape changing capability of $\mathbf{A}_{60^\circ,100^\circ}^{1~\textrm{m}}$. The second version possesses an identical shape changing capability and cytoskeletons that restrain module deformations beyond limit shapes. The third version possesses cytoskeletons that are partially stressed at both limit shapes. This is achieved by reducing the shape changing capability from $\mathbf{A}_{60^\circ,100^\circ}^{1~\textrm{m}}$ to $\mathbf{A}_{70^\circ,90^\circ}^{1~\textrm{m}}$. It should be noted that cytoskeleton lengths are computed for certain cell pressures that approximately result in desired central angles. Hence, their lengths vary slightly along an adaptive module due to edge effects. This is the reason why only average cytoskeleton lengths are given in Figure~\ref{pic:Figure_2_2}. The elastic energy of cellular structures can be neglected if frictionless cell corner hinges and stiff cell sides are assumed. Such a simplification minimizes influence factors and maximizes the verifiability of presented results. Therefore, frictionless cell corner hinges and a relatively large axial stiffness of $EA=1$~GN are assumed for both cell sides and cytoskeletons. It is important to note that these assumptions do not considerably affect the performance of the considered cellular structures.

                \begin{figure}[htbp]
                    \begin{center}
                        \subfloat[]{
                            \includegraphics[height=0.3\textwidth]{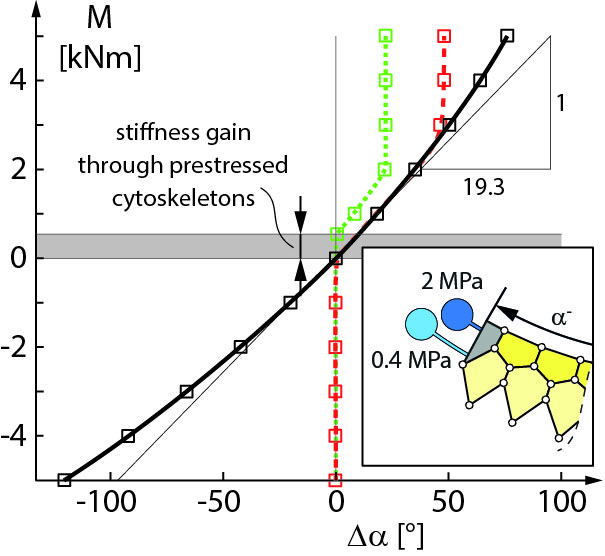}}\hspace{5mm}
                        \subfloat[]{
                            \includegraphics[height=0.3\textwidth]{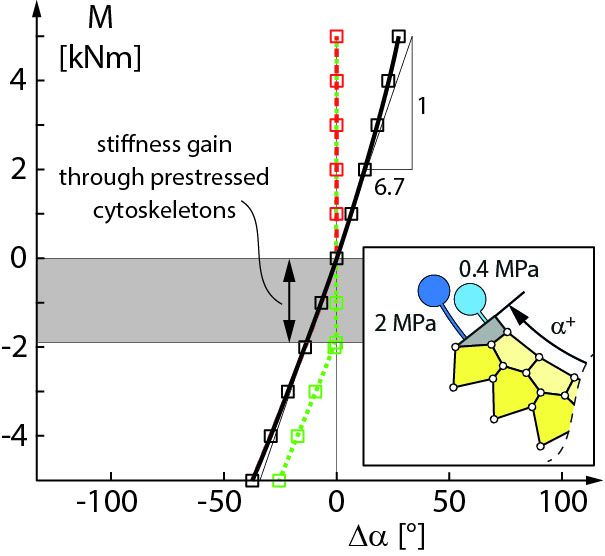}}\hspace{5mm}
                        \subfloat{
                            \includegraphics[height=0.3\textwidth]{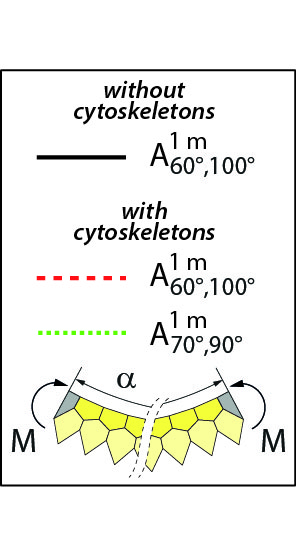}}
                        \caption{Change of central angle due to end moments at (a) lower and (b) upper limit shape for a one meter wide module.}
                        \label{pic:Figure_2_3}
                    \end{center}
                \end{figure}

                \noindent Central angle changes due to end moments at both limit shapes are shown in Figure~\ref{pic:Figure_2_3} for all three module versions. It can be seen that a module without cytoskeletons possesses, depending on cell pressures, a stiffness between $6.7-19.3^\circ$/kNm. This corresponds to the stiffness of a steel plate with a thickness between 5-8~mm. Introducing cytoskeletons restrains module deformations beyond limit shapes. Furthermore, the module stiffness at both limit shapes increases to $0.05-0.1^\circ$/kNm for loads between 0.5-2.0~kNm if cytoskeletons are partially prestressed. This corresponds to the stiffness of a steel plate with a thickness between 31-39~mm. It should be noted that the stiffness of cytoskeletons does not affect the stiffness gain shown in Figure~\ref{pic:Figure_2_3}. However, it does affect the tangent in that region. Furthermore, cytoskeletons are relatively thin and thus lightweight since their axial forces are well below the axial forces found in cell sides.


            \subsubsection{Weight}
\begin{figure}[htbp]
                    \begin{center}
                        \subfloat[]{
                            \includegraphics[height=0.07\textwidth]{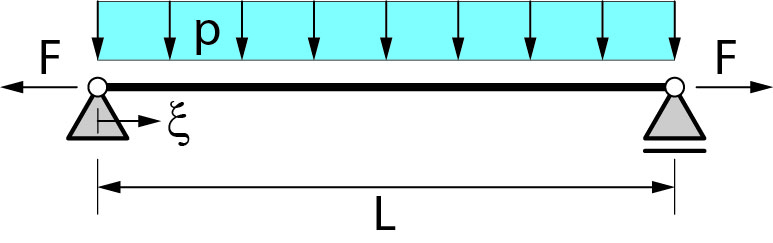}}\hspace{5mm}
                        \subfloat[]{
                            \includegraphics[height=0.07\textwidth]{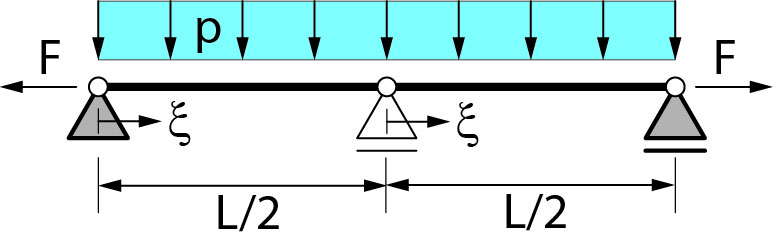}}\hspace{5mm}
                        \subfloat[]{
                            \includegraphics[height=0.07\textwidth]{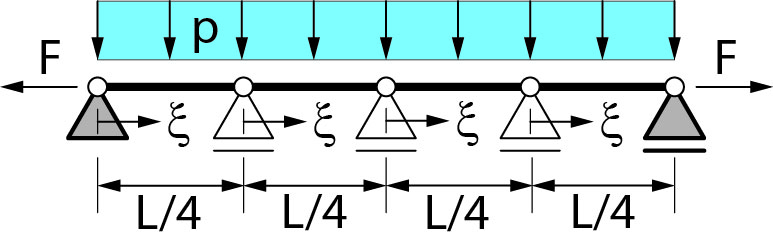}}\hspace{5mm}
                        \subfloat{
                            \includegraphics[height=0.07\textwidth]{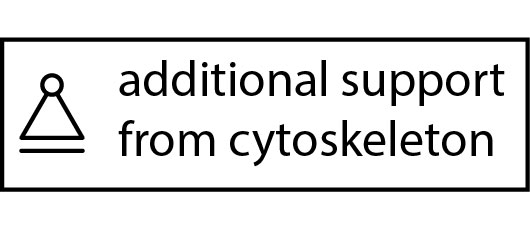}}                        \caption{Numerical models of a single cell side for increasingly refined cytoskeletons. (a) Initial configuration without cytoskeletons ($q=0$), (b) first ($q=1$) and (c) second ($q=2$) refinement level.}
                        \label{pic:Figure_2_4}
                    \end{center}
                \end{figure}

                The second kind of cytoskeletons reduces pressure induced bending moments in cell sides and thus the overall weight of a cellular structure. This is achieved by providing additional supports along a cell side as illustrated in Figure~\ref{pic:Figure_2_4}.

                \noindent The required thickness for a cell side that is subjected to an axial force $F$, a bending moment $M\left(\xi\right)$ and a yield strength $\sigma_y$ is

                \begin{align}
                    t\left(\xi\right)=\frac{\mid F \mid + \sqrt{F^2 + 24 \mid M\left(\xi\right) \mid \sigma_y}}{2 \sigma_y}\ \ \ \ \ \textrm{where}\ \ \ \ \ \xi \in\left[0,1\right].
                \end{align}

                \noindent The bending moment $M\left(\xi\right)$ along a cell side segment can be written as a function of the pressure load $p$ and the refinement level $q$ as

                \begin{align}
                    M\left(\xi,q\right) = 2^{-2q-1} p L^2 \left(1-\xi\right) \xi
                \end{align}

                \noindent so that the cell side thickness is

                \begin{align}
                    t\left(\xi,q\right)=\frac{\mid F \mid + \sqrt{F^2 + 3 \cdot 4^{1-q} \mid p \mid L^2 \left(1-\xi\right) \xi \sigma_y}}{2 \sigma_y}.
                \end{align}

                \noindent Consequently, the weight ratio $w_{q_2}/w_{q_1}$ of a single cell side for different refinement levels is bounded by $F=0$ and $p=0$ so that

                \begin{align}
                    2^{q_1-q_2} \leq w_{q_2}/w_{q_1}=t_{q_2}/t_{q_1} \leq 1\ \ \ \ \ \textrm{for}\ \ \ \ \ q_2 \geq q_1.
                \end{align}

\begin{figure}[htbp]
                    \begin{center}
                        \subfloat[]{
                            \includegraphics[height=0.32\textwidth]{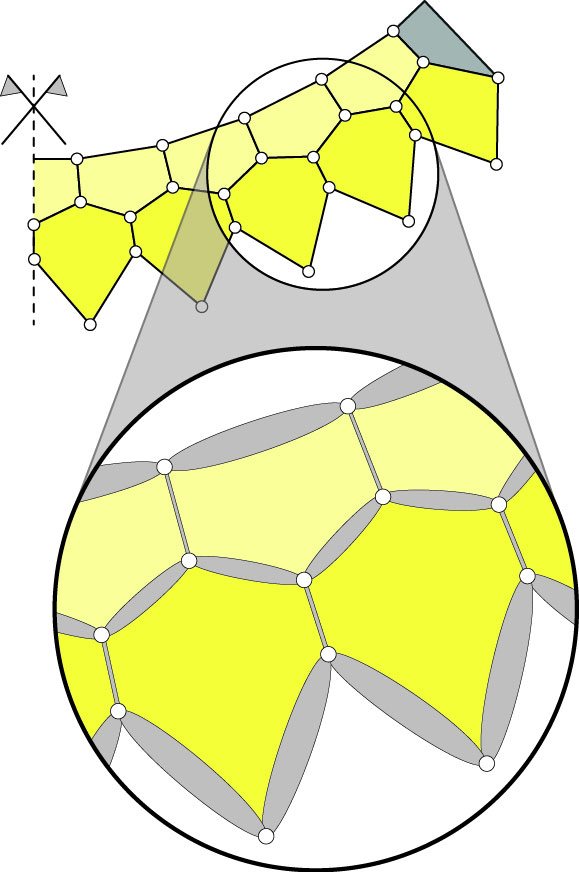}}\hspace{15mm}
                        \subfloat[]{
                            \includegraphics[height=0.32\textwidth]{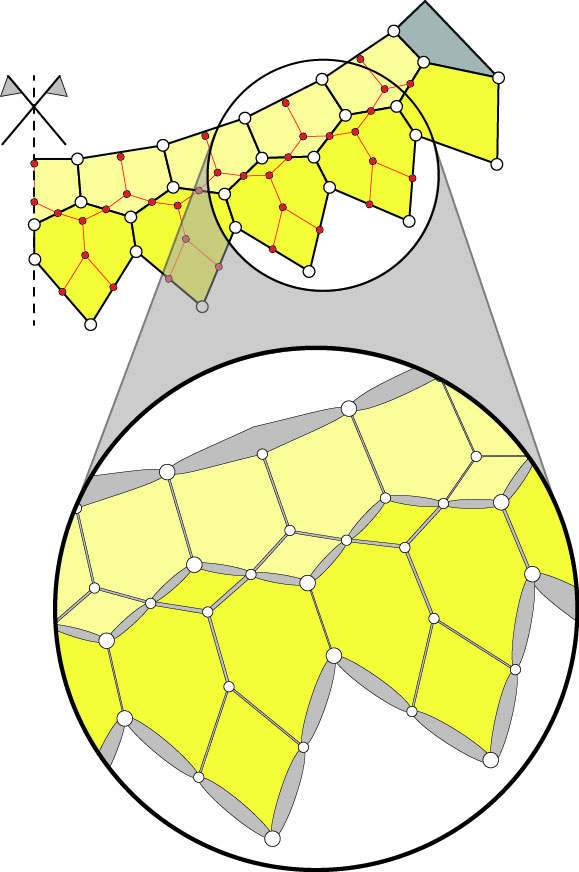}}\hspace{15mm}
                        \subfloat[]{
                            \includegraphics[height=0.32\textwidth]{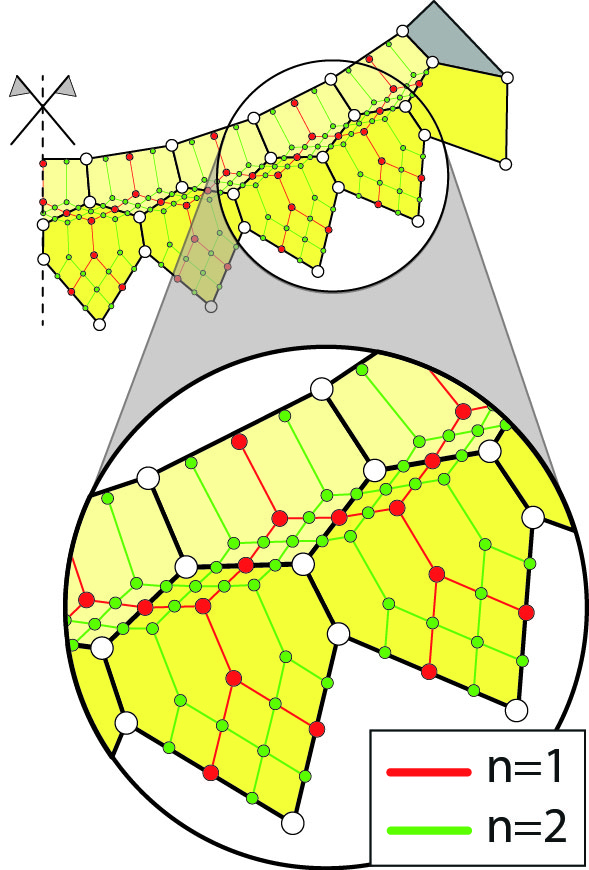}}
                        \caption{Cell side thicknesses of adaptive module $\mathbf{A}_{60^\circ,100^\circ}^{1~\textrm{m}}$ for refinement level (a) $q=0$ and (b) $q=1$. (c) Potential cytoskeletons for refinement level $q=2$.}
                        \label{pic:Figure_2_5}
                    \end{center}
                \end{figure}

                \noindent Two levels of cytoskeletons that provide load paths between pressure loaded cell sides are shown in Figure~\ref{pic:Figure_2_5}. It can be seen that cytoskeletons form four-bar linkages that do not restrain cell deformations.

                \noindent Possible weight reductions for various materials are summarized in Table~\ref{tab:Weight}. Note that a weight reduction\footnote{Weight of cytoskeletons is considered whereas weight of connectors is not considered. Furthermore, for the sake of simplicity, the first and last cell of the second row do not contain any cytoskeletons.} of up to 36\% is possible with a single level of cytoskeletons. More levels of cytoskeletons or the use of optimized cytoskeletons would further reduce the overall weight. Hence, the computed weight savings are a lower bound.

                \begin{table}[htbp]
                    \begin{center}
                        \begin{tabular}{l|cc|cc|c}
                                        & $\rho$        & $\sigma_y$ & $w_0$ & $w_1$ & $w_1/w_0$\\
                                        & [kg/m$^3$]    & [MPa]      & [kg]  & [kg]  & [-]\\\hline
                            PA6GF25     & 1,320         & 130        & 42.6  & 29.8  & 0.70\\
                            PEEKCF30    & 1,440         & 224        & 34.6  & 23.7  & 0.69\\
                            Ti10V2Fe3Al & 4,650         & 1,105      & 48.3  & 31.6  & 0.65\\
                            X2NiCoMo1895& 8,200         & 1,815      & 66.0  & 42.7  & 0.65
                        \end{tabular}
                        \caption{Weight of adaptive module $\mathbf{A}_{60^\circ,100^\circ}^{1~\textrm{m}}$ for different materials and two refinement levels. Note that a one meter wide module is assumed.}
                        \label{tab:Weight}
                    \end{center}
                \end{table}


        \subsection{Mechanical and Rigid Modules}
            The stiffness of an adaptive module is inversely proportional to its shape changing capability. Consequently, adaptive modules can not be used for regions of a structure that undergo large shape changes. Instead it is best to use mechanical modules $\mathbf{M}_{\alpha^-,\alpha^+}^L$ that are based on one of the construction principles shown in Figure~\ref{pic:Figure_1_2}a-d. On the other hand, rigid modules $\mathbf{R}_{\pm \alpha}^L$ can be used for regions that experience only minor shape changes.


        \subsection{Assembly of Modules}
\begin{figure}[htbp]
                \begin{center}
                    \includegraphics[height=0.45\textwidth]{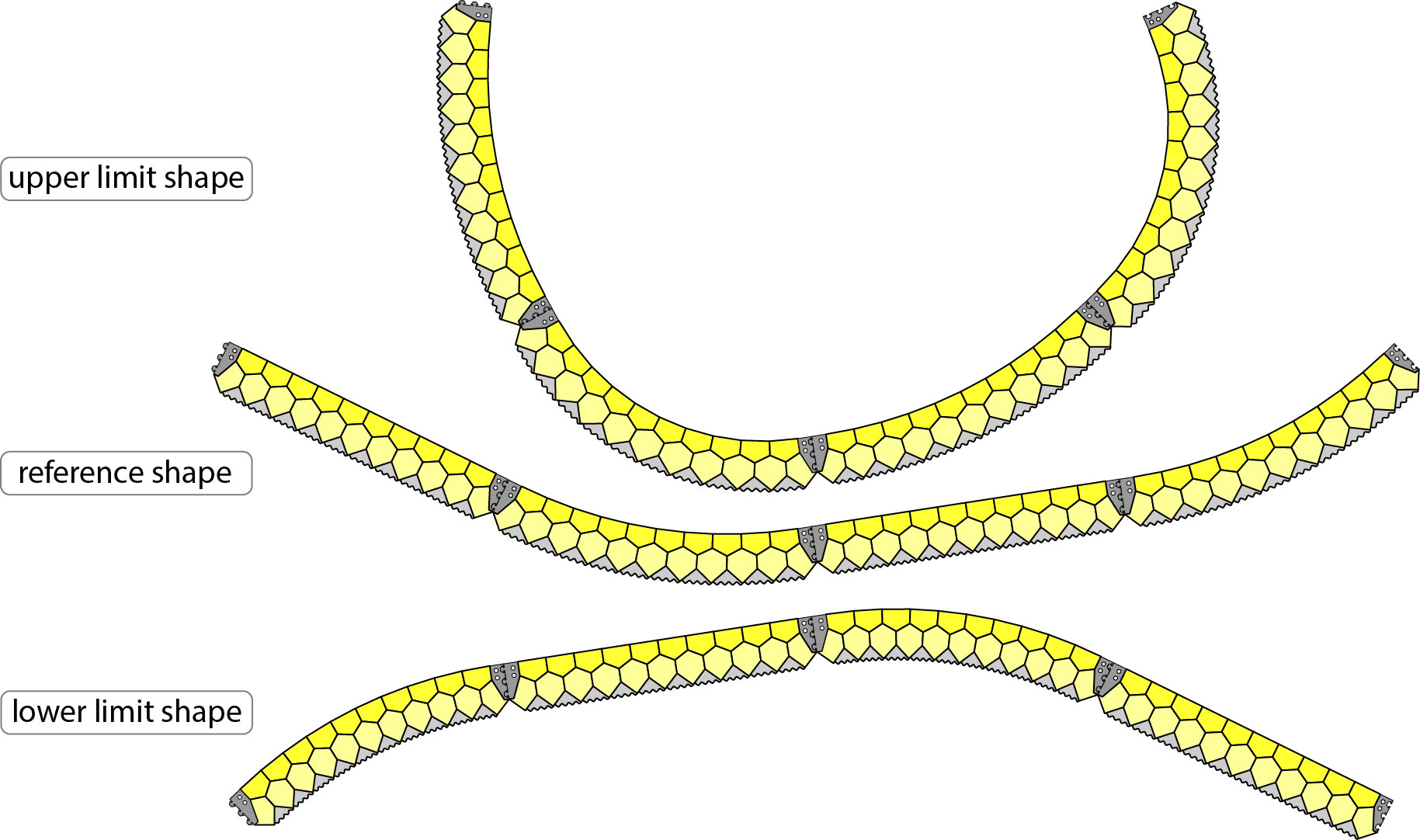}
                    \caption{Reference and limit shapes of four connected modules.}
                    \label{pic:Figure_2_6}
                \end{center}
            \end{figure}

            A connection between two modules is indicated by the symbol ``$\Box$'' so that, for example,

            \begin{align}
                \mathbf{S}=\left[ \mathbf{A}_{\mp 35^\circ}^L \ \Box\  \mathbf{A}_{0^\circ,70^\circ}^L \ \Box\  \mathbf{A}_{\mp 35^\circ}^L \ \Box\  \mathbf{A}_{0^\circ,70^\circ}^L \right]
            \end{align}

            \noindent is a structure that is made from four adaptive modules. The reference\footnote{The reference shape of an adaptive module is assumed to be the average of its limit shapes.} and limit shapes of this structure are shown in Figure~\ref{pic:Figure_2_6}. It should be noted that cell row pressures of each module have to be controlled separately in order to fully exploit the shape changing capability of a modular structure. This is in contrast to a purpose made structure where all cell geometries are tailored for a given set of targets shapes and cell pressures.

            \noindent A consequence of the separate module pressurization is that a single adaptive module can be considered to be a subset of two smaller modules

            \begin{align}
                \mathbf{A}_{\alpha^-,\alpha^+}^{L} \subset \left[ \mathbf{A}_{\alpha_1^-,\alpha_1^+}^{L_1} \ \Box\  \mathbf{A}_{\alpha_2^-,\alpha_2^+}^{L_2} \right]\ \ \ \ \ \textrm{or}\ \ \ \ \
                \mathbf{M}_{\alpha^-,\alpha^+}^{L} \subset \left[ \mathbf{M}_{\alpha_1^-,\alpha_1^+}^{L_1} \ \Box\  \mathbf{M}_{\alpha_2^-,\alpha_2^+}^{L_2} \right]
            \end{align}

            \noindent if

            \begin{align}
                L=L_1+L_2,\ \ \ \ \ \frac{\alpha^-}{L} \geq \left\{\frac{\alpha_1^-}{L_1}, \frac{\alpha_2^-}{L_2}\right\}\ \ \ \ \ \textrm{and}\ \ \ \ \ \frac{\alpha^+}{L} \leq \left\{\frac{\alpha_1^+}{L_1}, \frac{\alpha_2^+}{L_2}\right\}.
            \end{align}

            \noindent A similar relationship exists for rigid modules since

            \begin{align}
                \mathbf{R}_{\alpha}^{L} = \left[ \mathbf{R}_{\alpha L_1/L}^{L_1} \ \Box\  \mathbf{R}_{\alpha L_2/L}^{L_2} \right]\ \ \ \ \ \textrm{if}\ \ \ \ \ L=L_1+L_2.
            \end{align}


    \section{Discretization of Target Shapes}
\begin{figure}
            \begin{minipage}[c]{0.5\textwidth}
                \begin{center}
                    \subfloat[]{
                        \includegraphics[width=0.6\textwidth]{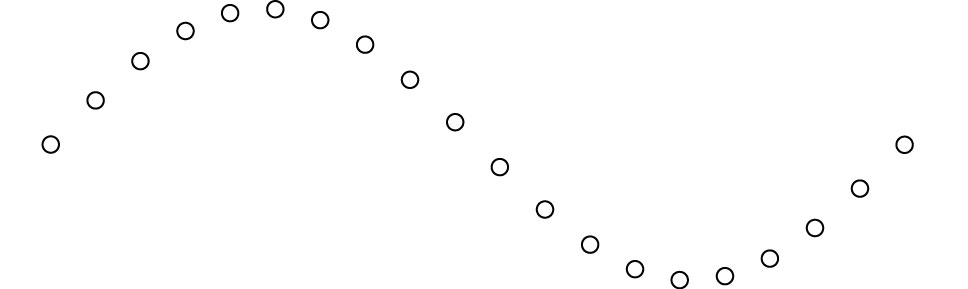}}

                    \subfloat[]{
                        \includegraphics[width=0.6\textwidth]{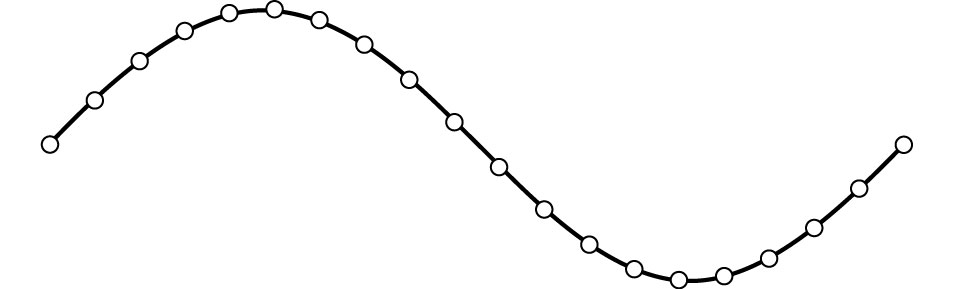}}

                    \subfloat[]{
                        \includegraphics[width=0.6\textwidth]{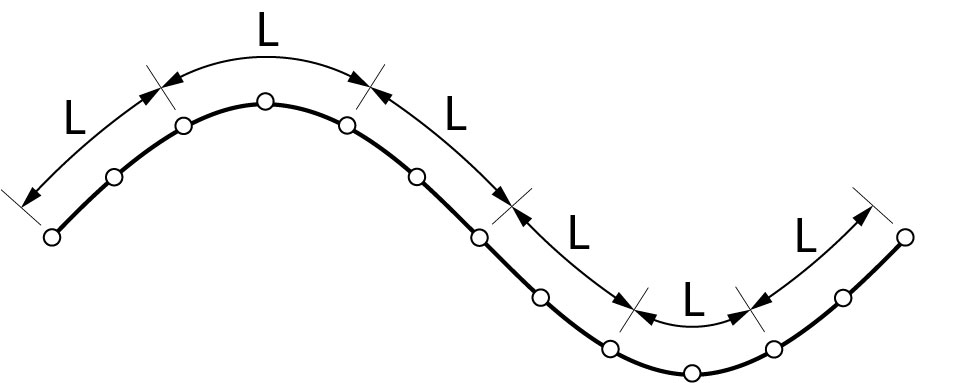}}
                \end{center}
            \end{minipage}
            \begin{minipage}[c]{0.5\textwidth}
                \begin{center}
                    \subfloat[]{
                        \includegraphics[width=0.8\textwidth]{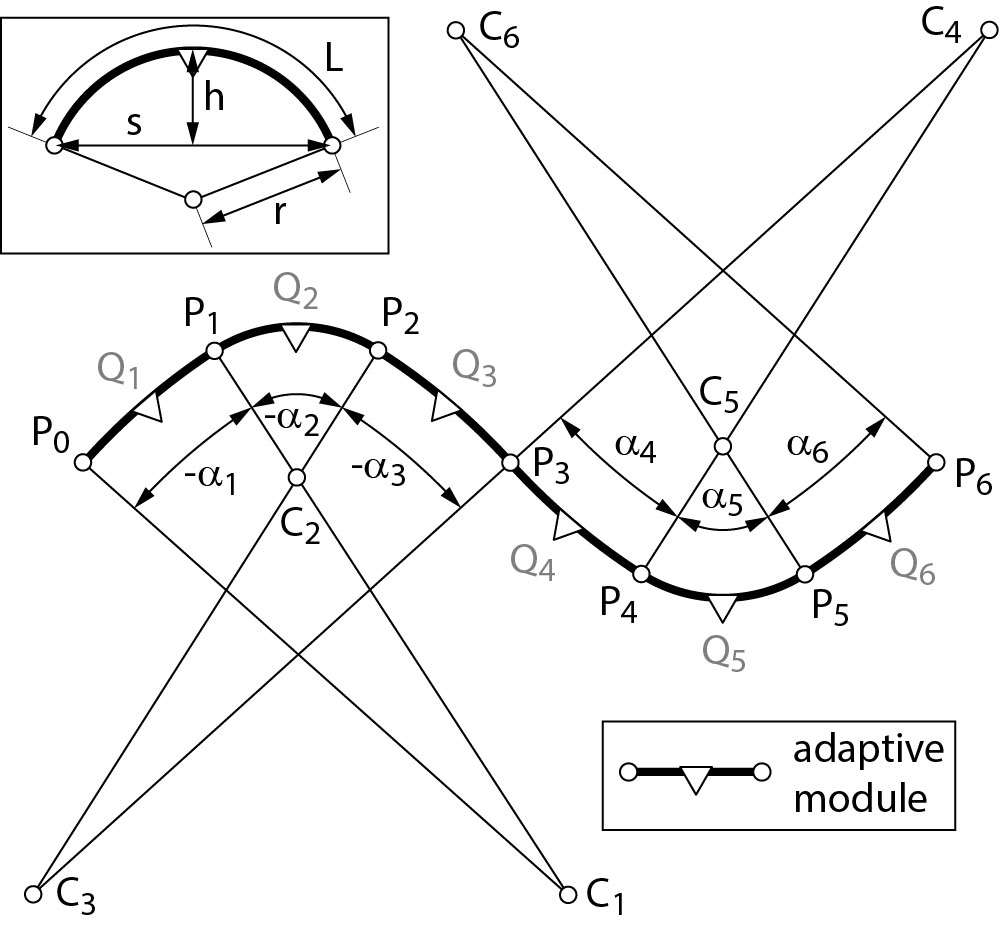}}
                \end{center}
            \end{minipage}
            \caption{Interpolation of target shapes with a set of different adaptive modules. (a) Target shapes are defined by randomly spaced points; (b) Coordinates of points are interpolated with polynomials; (c) Interpolated curve is split into $2m$ line segments with an arc length $L/2$; (d) Line segments are interpolated with a set of $n$ different modules. Note that $C^1$ continuity between modules is enforced.}
            \label{pic:Figure_3_1}
        \end{figure}

        All continuous target shapes can be accurately represented by infinitesimally small circular arcs. The error introduced by circular arcs with finite, constant arc lengths is usually small so that standardized modules can be used instead. This section introduces an algorithm that interpolates target shapes with a given set of $n$ different modules $\left\{\mathbf{A}_{\alpha_1^-, \alpha_1^+}^{L},\ldots,\mathbf{A}_{\alpha_n^-, \alpha_n^+}^{L}\right\}$ such that the approximation error is minimal. Furthermore, $C^1$ continuity between modules is enforced. The functionality of this algorithm and the used notation is summarized in Figure~\ref{pic:Figure_3_1}.\newline

        The chord length $s$ of a circular arc with an arc length $L$ and central angle $\alpha$ is

        \begin{align}
            s = \frac{2 L}{\alpha} \sin\left(\frac{\alpha}{2}\right)
        \end{align}

        \noindent so that the first and second derivatives with respect to $\alpha$ are

        \begin{align}
            \frac{\partial s}{\partial \alpha} = \frac{1}{\alpha} \left( L \cos \left( \frac{\alpha}{2} \right) - s \right)\ \ \ \ \ \textrm{and}\ \ \ \ \
            \frac{\partial^2 s}{\partial \alpha^2} = - \frac{1}{4} \left( \frac{8}{\alpha} \frac{\partial s}{\partial \alpha} + s \right)\ \ \ \ \ \textrm{since}\ \ \ \ \
            r = \frac{L}{\alpha}.
        \end{align}

        \noindent The height $h$ of a circular arc is

        \begin{align}
            h = \frac{L}{\alpha} \left( 1 - \cos\left( \frac{\alpha}{2} \right) \right)
        \end{align}

        \noindent and its first and second derivatives with respect to $\alpha$ are

        \begin{align}
            \frac{\partial h}{\partial \alpha} = \frac{1}{\alpha} \left( \frac{L}{2} \sin \left( \frac{\alpha}{2} \right) - h \right)\ \ \ \ \ \textrm{and}\ \ \ \ \
            \frac{\partial^2 h}{\partial \alpha^2} = -\frac{1}{4} \left( \frac{8}{\alpha} \frac{\partial h}{\partial \alpha} + h - \frac{L}{\alpha} \right).
        \end{align}

        \noindent The nodal $\mathbf{P}_j$ and central $\mathbf{Q}_j$ coordinates of circular arcs with an enforced $C^1$ intersegment continuity are

        \begin{align}
            \mathbf{P}_j = \mathbf{P}_{j-1} +
            \left[
            \begin{array}{r}
                \cos\left( \theta_j \right)\\
                \sin\left( \theta_j \right)
            \end{array}
            \right] s_j\ \ \ \ \ \textrm{and}\ \ \ \ \
            \mathbf{Q}_j = \mathbf{P}_{j-1}
            + \frac{1}{2}
            \left[
            \begin{array}{r}
                \cos\left( \theta_j \right)\\
                \sin\left( \theta_j \right)
            \end{array}
            \right] s_j
            -
            \left[
            \begin{array}{r}
                -\sin\left( \theta_j \right)\\
                 \cos\left( \theta_j \right)
            \end{array}
            \right] h_j
        \end{align}

        \noindent where the angle $\theta_j$ is

        \begin{align}
            \theta_j = \frac{1}{2} \sum_{i=0}^{j-1} \left( \alpha_i+\alpha_{1+i} \right).
        \end{align}

        \noindent Note that $\alpha_0$ defines the gradient of the modular structure at $\mathbf{P}_0$. The first derivatives of $\mathbf{P}_j$ and $\mathbf{Q}_j$ with respect to $\boldsymbol{\alpha}$ are

        \begin{align}
            \frac{\partial \mathbf{P}_j^T}{\partial \boldsymbol{\alpha}} = \frac{\partial \mathbf{P}_{j-1}^T}{\partial \boldsymbol{\alpha}}
            + \frac{1}{2}
            \left[
            \begin{array}{r}
                -\sin\left( \theta_j \right)\\
                 \cos\left( \theta_j \right)
            \end{array}
            \right]^T s_j \mathbf{v}_1
            +
            \left[
            \begin{array}{c}
                \cos\left( \theta_j \right)\\
                \sin\left( \theta_j \right)
            \end{array}
            \right]^T \frac{\partial s_j}{\partial \alpha_j} \mathbf{v}_2
        \end{align}

        \noindent and

        \begin{align}
            \frac{\partial \mathbf{Q}_j^T}{\partial \boldsymbol{\alpha}} = \frac{1}{2} \left( \frac{\partial \mathbf{P}_{j-1}^T}{\partial \boldsymbol{\alpha}} + \frac{\partial \mathbf{P}_{j}^T}{\partial \boldsymbol{\alpha}} \right)
            + \frac{1}{2}
            \left[
            \begin{array}{c}
                \cos\left( \theta_j \right)\\
                \sin\left( \theta_j \right)
            \end{array}
            \right]^T h_j \mathbf{v}_1
            -
            \left[
            \begin{array}{r}
                -\sin\left( \theta_j \right)\\
                 \cos\left( \theta_j \right)
            \end{array}
            \right]^T \frac{\partial h_j}{\partial \alpha_j} \mathbf{v}_2.
        \end{align}

        \noindent Vectors $\mathbf{v}_1$, $\mathbf{v}_2$ and $\boldsymbol{\alpha}$ contain $m+1$ elements. The number of non-zero entries in $\mathbf{v}_1$ and $\mathbf{v}_2$ is $j+1$ so that

        \begin{align}
            \mathbf{v}_1 =
            \left[
            \begin{array}{ccccc|c}
                1 & 2 & \cdots & 2 & 1 & \mathbf{0}
            \end{array}
            \right]^T,\ \ \ \ \
            \mathbf{v}_2 =
            \left[
            \begin{array}{ccccc|c}
                0 & 0 & \cdots & 0 & 1 & \mathbf{0}
            \end{array}
            \right]^T\ \ \ \ \ \textrm{and}\ \ \ \ \
            \boldsymbol{\alpha} =
            \left[
            \begin{array}{ccc}
                \alpha_0 & \hdots & \alpha_m
            \end{array}
            \right]^T.
        \end{align}

        \noindent The second derivatives of $\mathbf{P}_j$ and $\mathbf{Q}_j$ with respect to $\boldsymbol{\alpha}$ are

        \begin{align}
            \frac{\partial^2 \mathbf{P}_j^T}{\partial \boldsymbol{\alpha}^2} = \frac{\partial^2 \mathbf{P}_{j-1}^T}{\partial \boldsymbol{\alpha}^2} -
            \left[
            \begin{array}{c}
                \cos \left( \theta_j \right)\\
                \sin \left( \theta_j \right)
            \end{array}
            \right]^T
            \left( \frac{1}{4} s_j \mathbf{v}_1 \mathbf{v}_1^T - \frac{\partial^2 s_j}{\partial \alpha_j^2} \mathbf{v}_2 \mathbf{v}_2^T \right) + \frac{1}{2}
            \left[
            \begin{array}{r}
                -\sin \left( \theta_j \right)\\
                 \cos \left( \theta_j \right)
            \end{array}
            \right]^T
            \frac{\partial s_j}{\partial \alpha_j} \left( \mathbf{v}_1 \mathbf{v}_2^T + \mathbf{v}_2 \mathbf{v}_1^T \right)
        \end{align}

        \noindent and

        \begin{align}
            \frac{\partial^2 \mathbf{Q}_j^T}{\partial \boldsymbol{\alpha}^2} = \frac{1}{2} \left( \frac{\partial^2 \mathbf{P}_{j-1}^T}{\partial \boldsymbol{\alpha}^2} + \frac{\partial^2 \mathbf{P}_{j}^T}{\partial \boldsymbol{\alpha}^2} \right) +
            \left[
            \begin{array}{r}
                -\sin \left( \theta_j \right)\\
                 \cos \left( \theta_j \right)
            \end{array}
            \right]^T
            \left( \frac{1}{4} h_j \mathbf{v}_1 \mathbf{v}_1^T - \frac{\partial^2 h_j}{\partial \alpha_j^2} \mathbf{v}_2 \mathbf{v}_2^T \right) +
            \frac{1}{2}
            \left[
            \begin{array}{r}
                \cos \left( \theta_j \right)\\
                \sin \left( \theta_j \right)
            \end{array}
            \right]
            \frac{\partial h_j}{\partial \alpha_j} \left( \mathbf{v}_1 \mathbf{v}_2^T + \mathbf{v}_2 \mathbf{v}_1^T \right).
        \end{align}

        \noindent The error norm for a least square interpolation of a set of points $\left\{ \mathbf{P}^0_1,\ldots,\mathbf{P}^0_m \right\}$ is

        \begin{align}
            \Pi = \sum_{j=1}^{m} \left( \Delta \mathbf{P}_j^T \Delta \mathbf{P}_j + {\Delta \mathbf{Q}_j^T} \Delta \mathbf{Q}_j \right) = \textrm{min.}
            \ \ \ \ \ \textrm{where, for example,}\ \ \ \ \ \Delta \mathbf{P}_j = \mathbf{P}_j - \mathbf{P}_j^0
        \end{align}

        \noindent so that the first derivatives with respect to central angles $\boldsymbol{\alpha}$ are

        \begin{align}
            \mathbf{f} = \frac{\partial \Pi}{\partial \boldsymbol{\alpha}} =
            2 \sum_{j=1}^{m} \left( {\frac{\partial \mathbf{P}_j^T}{\partial \boldsymbol{\alpha}}} \Delta \mathbf{P}_j + {\frac{\partial \mathbf{Q}_j^T}{\partial \boldsymbol{\alpha}}} \Delta \mathbf{Q}_j \right)
        \end{align}

        \noindent and the corresponding second derivatives are

        \begin{align}
            \mathbf{K} = \frac{\partial^2 \Pi}{\partial \boldsymbol{\alpha}^2} =
            2 \sum_{j=1}^{m} \left(
            \frac{\partial^2 \mathbf{P}_j^T}{\partial \boldsymbol{\alpha}^2} \Delta \mathbf{P}_j + \frac{\partial^2 \mathbf{Q}_j^T}{\partial \boldsymbol{\alpha}^2} \Delta \mathbf{Q}_j
            + {\frac{\partial \mathbf{P}_j^T}{\partial \boldsymbol{\alpha}}} \frac{\partial \mathbf{P}_j}{\partial \boldsymbol{\alpha}^T} + {\frac{\partial \mathbf{Q}_j^T}{\partial \boldsymbol{\alpha}}} \frac{\partial \mathbf{Q}_j}{\partial \boldsymbol{\alpha}^T}
            \right)
        \end{align}

        \noindent since

        \begin{align}
            \mathbf{P}_0 = \mathbf{P}_0^0\ \ \ \ \ \textrm{and therefore}\ \ \ \ \ \frac{\partial \mathbf{P}_0}{\partial \boldsymbol{\alpha}} = \mathbf{0}.
        \end{align}

        \noindent Based on a Newton-Raphson scheme, an update for the angles $\boldsymbol{\alpha}$ is

        \begin{align}
            \Delta \boldsymbol{\alpha} = -\mathbf{K}^{-1} \mathbf{f}.
        \end{align}

        \noindent The upper and lower bounds for each angle $\alpha_j$ are defined by the chosen module $\mathbf{A}_{\alpha_j^-, \alpha_j^+}^L$. These constraints are iteratively enforced with Lagrange multipliers. Finally, the circle center $\mathbf{C}_j$ can be computed by using the following equalities

        \begin{align}
            \mid \mathbf{P}_{j-1} - \mathbf{C}_j \mid = \mid \mathbf{P}_{j} - \mathbf{C}_j \mid = \mid \mathbf{Q}_{j} - \mathbf{C}_j \mid
        \end{align}

        \noindent so that

        \begin{align}
            \mathbf{C}_j = \mathbf{Q}_j + \frac{1}{2} \left( \mathbf{P}_{j-1} + \mathbf{P}_{j} - 2 \mathbf{Q}_{j} \right) \frac{r_j}{h_j}.
        \end{align}


    \section{Adaptive Structures}

        \subsection{Airfoil}
\begin{figure}[htbp]
                \begin{center}
                    \subfloat[]{
                        \label{pic:Figure_4_1a}
                        \includegraphics[width=0.75\textwidth]{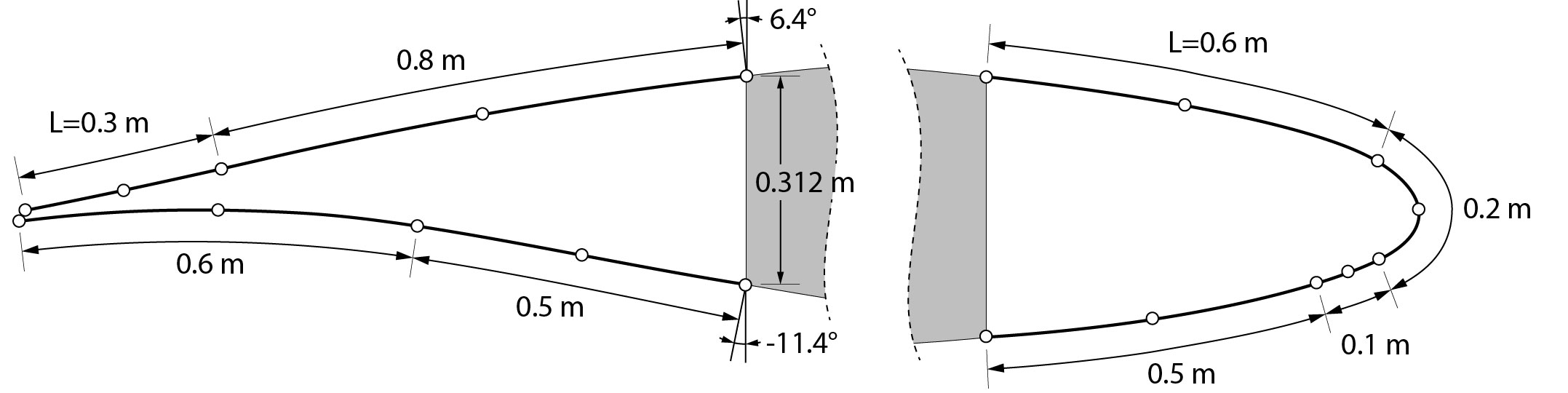}}

                    \subfloat[]{
                        \label{pic:Figure_4_1b}
                        \includegraphics[width=0.75\textwidth]{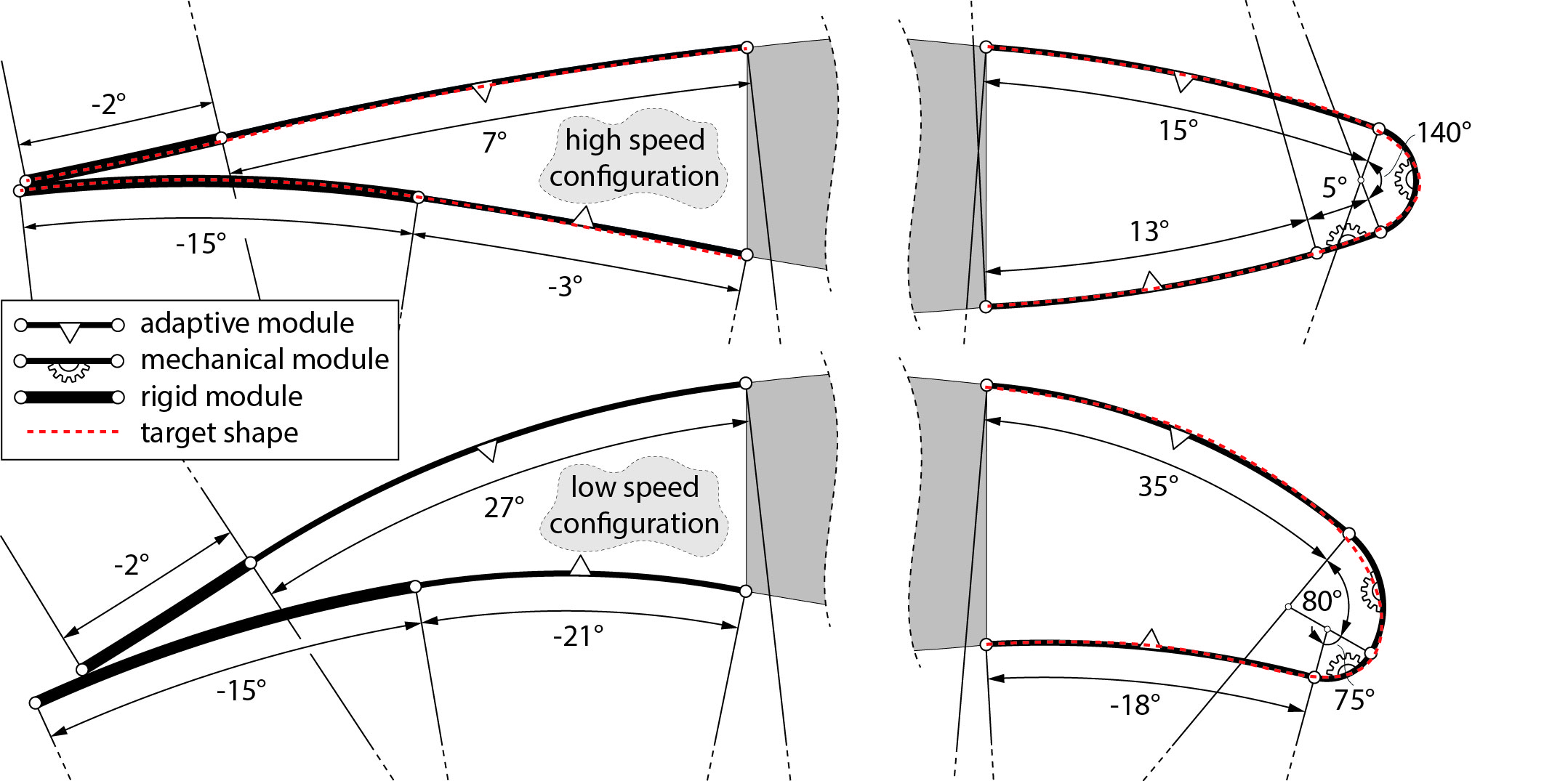}}

                    \subfloat[]{
                        \label{pic:Figure_4_1c}
                        \includegraphics[width=0.75\textwidth]{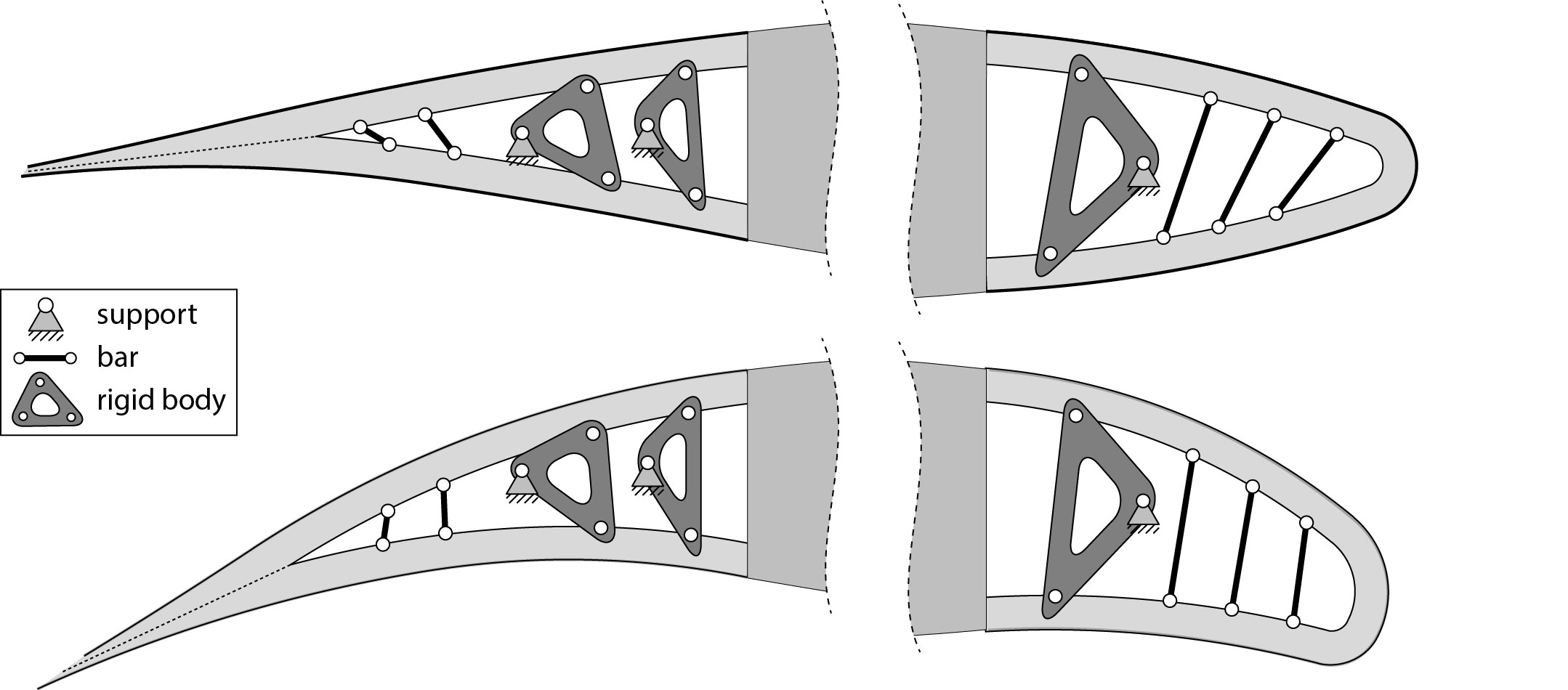}}
                    \caption{(a) Interpolated curves are split into line segments with given arc lengths (shown for the first target shapes); (b) Line segments are interpolated with adaptive, rigid modules. Note that upper side of trailing edge is assumed to change its inclination by $\Delta\alpha=20^\circ$; (c) Potential mechanisms for module thicknesses $t=50$~mm.}
                \end{center}
            \end{figure}

            An aircraft wing is the first example that demonstrates the approximation power of adaptive modules. Points that describe the geometry of an optimized leading, trailing edge during high, low flight speeds are summarized in Appendix~\ref{sec:CoordinatesAirfoil}\footnote{Only a single set of points is given for the trailing edge since available data is based on a rigid Fowler flap. Nevertheless, an adaptive version of the trailing edge that can change the inclination of the upper surface by $20^\circ$ is considered.}. Closed form expressions for the geometry of the leading, trailing edges are obtained by polynomial interpolation. The resulting curves are then split into a number of segments with given arc lengths as shown in Figure~\ref{pic:Figure_4_1a}. Note that a point at the center of each segment is computed for a subsequent interpolation with circular arcs, modules. The optimal number of segments and their arc lengths is found by a combinatorial optimization approach\footnote{Segment arc lengths need to have a common divisor so that standardized modules can be used. Furthermore, the smallest possible number of segments that results in an acceptable interpolation error is used.}. This is possible since the underlying problem, total number of modules is usually small. A $C^1$ continuous interpolation of the target shapes with circular arcs, modules is shown in Figure~\ref{pic:Figure_4_1b}. It can be seen that all target shapes are accurately represented by a small number of modules with a common divisor of $L=0.1$~m. The segments with an arc length of $0.1$~m and $0.2$~m experience relatively large shape changes. Hence it is best to use mechanical modules for these segments. The resulting combination of adaptive, mechanical and rigid modules for the leading edge is

            \begin{align}
                \mathbf{S} = \left[
                \mathbf{A}_{15^\circ,35^\circ}^{0.6~\textrm{m}} \ \Box\
                \mathbf{M}_{80^\circ,140^\circ}^{0.2~\textrm{m}} \ \Box\
                \mathbf{M}_{5^\circ,75^\circ}^{0.1~\textrm{m}} \ \Box\
                \mathbf{A}_{-18^\circ,13^\circ}^{0.5~\textrm{m}} \right]
            \end{align}

            \noindent and for the trailing edge is

            \begin{align}
                \mathbf{S} =
                \begin{cases}
                    \left[
                    \mathbf{R}_{-2^\circ}^{0.3~\textrm{m}} \ \Box\
                    \mathbf{A}_{7^\circ,27^\circ}^{0.8~\textrm{m}} \right] & \mbox{top surface}\\
                    \left[
                    \mathbf{R}_{-15^\circ}^{0.6~\textrm{m}} \ \Box\
                    \mathbf{A}_{-21^\circ,-3^\circ}^{0.5~\textrm{m}} \right] & \mbox{bottom surface.}
                \end{cases}
            \end{align}

            \noindent The stiffness of an adaptive airfoil can be increased by using internal mechanisms, Figure~\ref{pic:Figure_4_1c}. Potential mechanisms can be found by computing, for all possible airfoil shapes, equidistant pairs of points along the inner sides of a leading, trailing edge. Another equidistant, fixed point inside the leading, trailing edge can be found for some point pairs. These additional points can be used as supports for rigid body mechanisms since their position is invariant.

        \subsection{Passenger Seat}
\begin{figure}
                \begin{minipage}[c]{0.5\textwidth}
                    \begin{center}
                        \subfloat[]{
                            \label{pic:Figure_4_2a}
                            \includegraphics[height=1.5\textwidth]{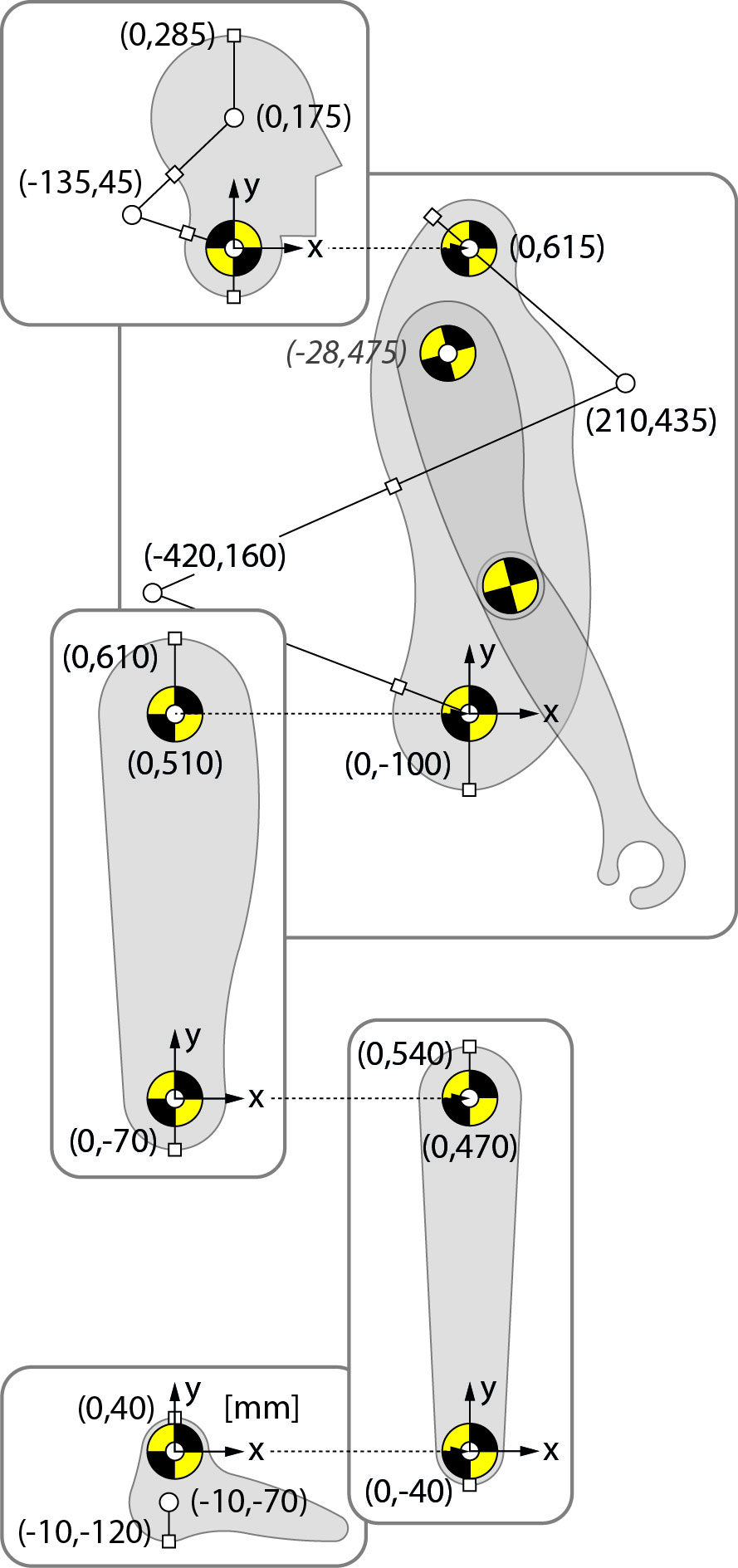}}
                    \end{center}
                \end{minipage}
                \begin{minipage}[c]{0.5\textwidth}
                    \begin{center}
                        \subfloat[]{
                            \label{pic:Figure_4_2b}
                            \includegraphics[height=1.5\textwidth]{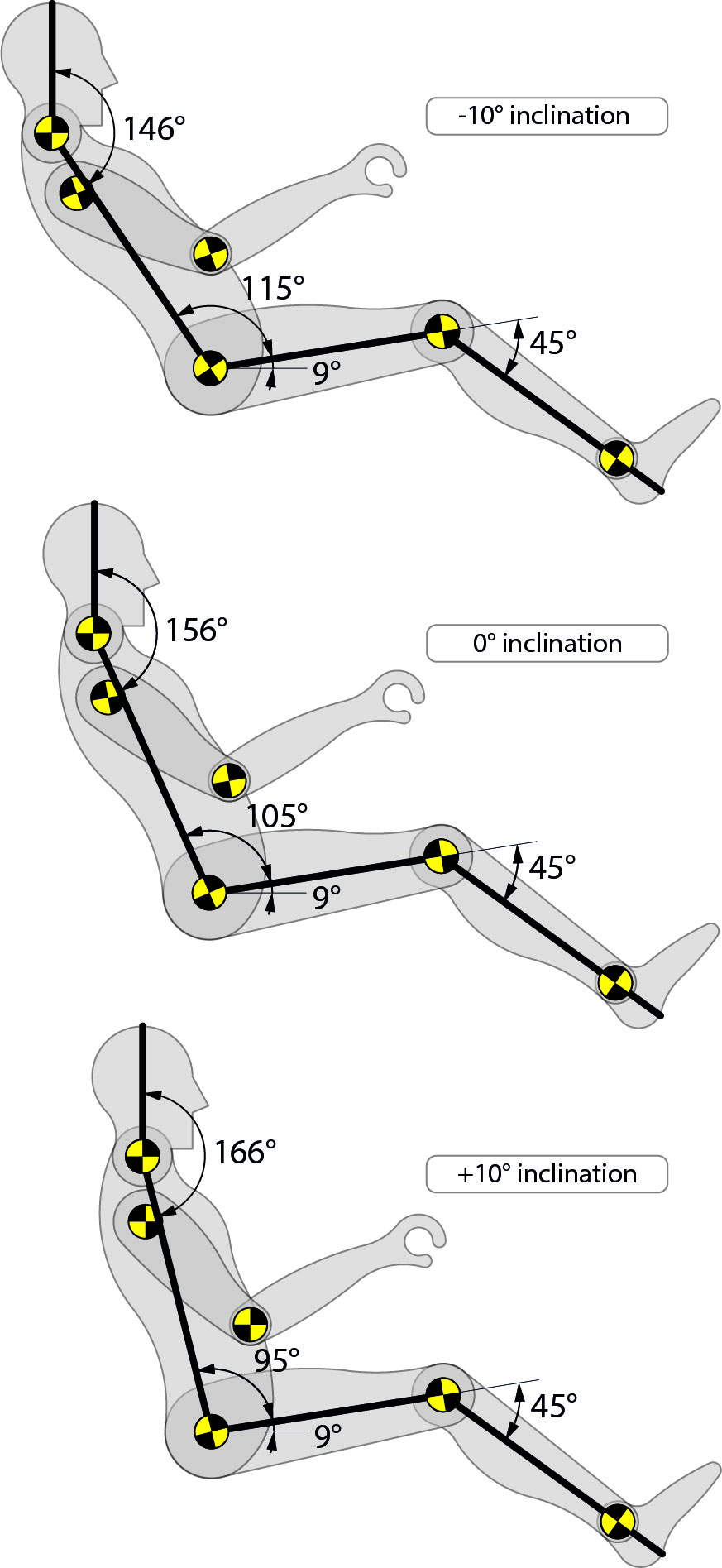}}
                    \end{center}
                \end{minipage}
                \caption{(a) Geometry and mechanism of a two meter tall person; (b) Sitting position for different backrest inclinations.}
            \end{figure}

            The second example is a passenger seat that can change its inclination by $\pm 10^\circ$ and adapt to persons with heights between $1.7-1.9$~m. The geometry of a $2$~m tall person is defined in Figure~\ref{pic:Figure_4_2a} by a sequence of $C^1$ continuous circular arcs\footnote{The geometric representation of the human body with circular arcs is not linked to an interpolation with modules. If possible it is best to directly use closed form expressions for the geometry rather than a number of discrete points.}. Data for different body heights is obtained by scaling. Varying seat inclinations are modeled with a rigid body rotation of the torso around the hip and a corresponding displacement of the head, Figure~\ref{pic:Figure_4_2b}.

\begin{figure}[htbp]
                \begin{center}
                    \subfloat[]{
                        \label{pic:Figure_4_3a}
                        \includegraphics[width=0.8\textwidth]{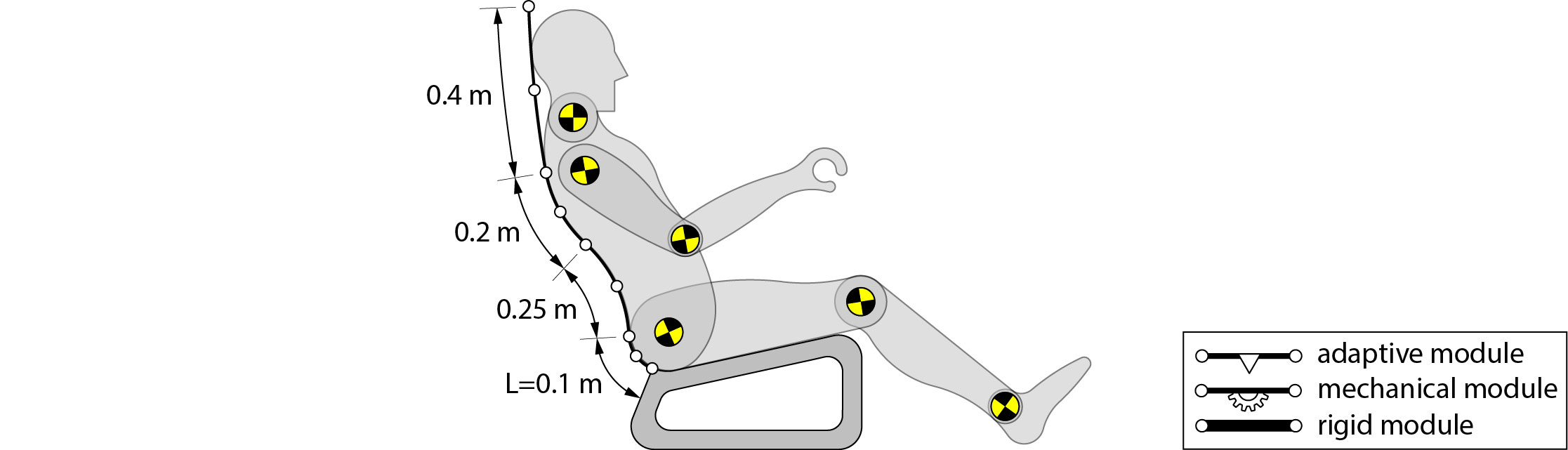}}

                    \subfloat[]{
                        \label{pic:Figure_4_3b}
                        \includegraphics[width=0.8\textwidth]{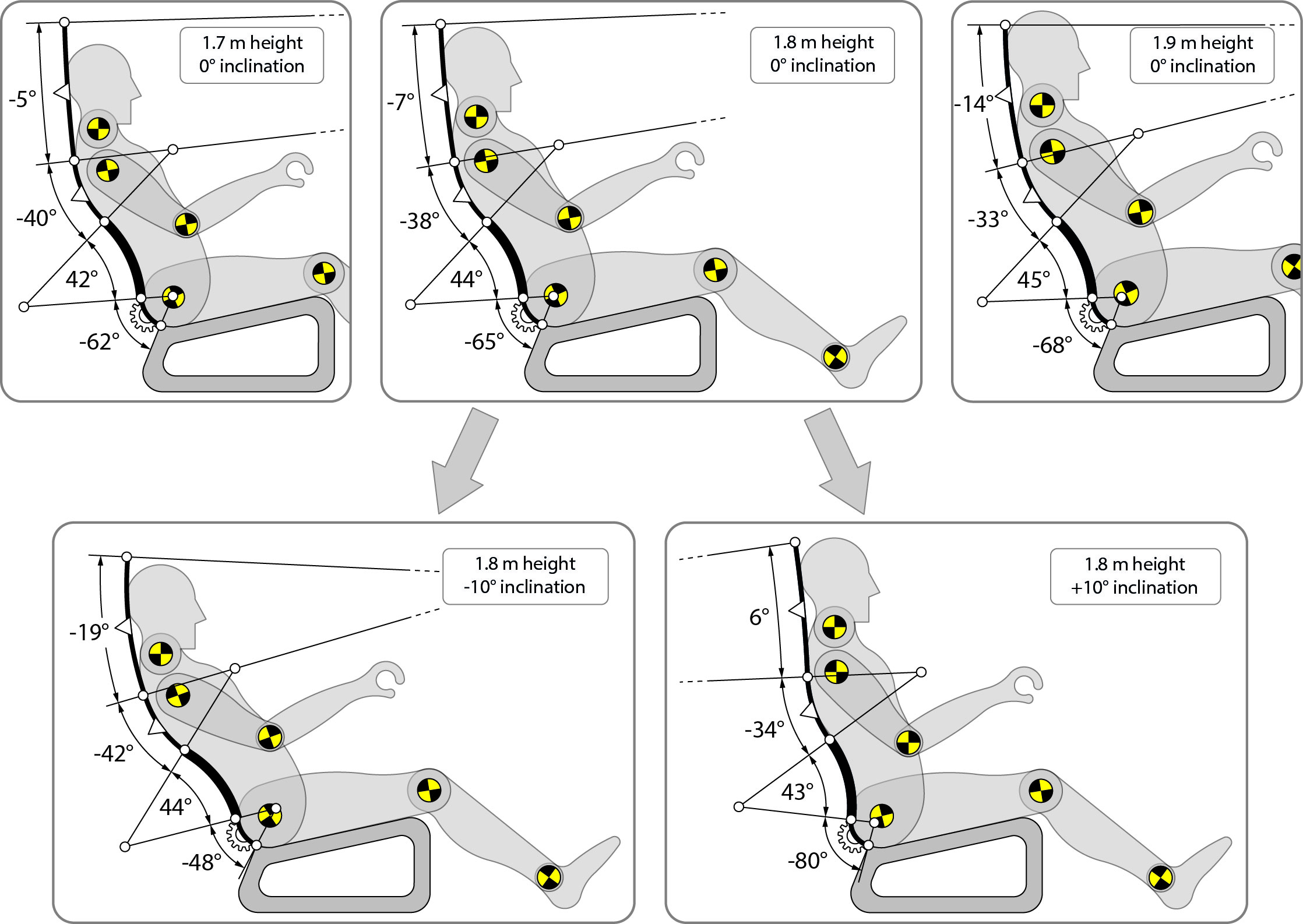}}
                    \caption{(a) Backrest is split into four segments with given arc lengths (superimposed person is $1.8$~m tall, standard sitting position); (b) Seat can change its inclination by $\pm 10^\circ$ and adapt to persons with heights between $1.7-1.9$~m.}
                \end{center}
            \end{figure}

            \noindent The backrest is assumed to consist of four modules as shown in Figure~\ref{pic:Figure_4_3a} for a standard sitting position and a superimposed person with a height of $1.8$~m. Note that all segment lengths are a multiple of $L=0.05$~m. Interpolations with four adaptive modules for varying body heights and seat inclinations are shown in Figure~\ref{pic:Figure_4_3b}. It can be seen that the corresponding approximation errors are acceptably small for all five target shapes. The subtended angle of the $0.25$~m segment varies only slightly so that it can be considered to be rigid. On the other hand, the $0.1$~m segment experiences relatively large shape changes. Hence it is best to use a mechanical module for this segment. The combination of adaptive, mechanical and rigid modules for a passenger seat is

            \begin{align}
                \mathbf{S} = \left[
                \mathbf{A}_{-19^\circ,6^\circ}^{0.4~\textrm{m}} \ \Box\
                \mathbf{A}_{-42^\circ,-33^\circ}^{0.2~\textrm{m}} \ \Box\
                \mathbf{R}_{44^\circ}^{0.25~\textrm{m}} \ \Box\
                \mathbf{M}_{-80^\circ,-48^\circ}^{0.1~\textrm{m}} \right].
            \end{align}

        \subsection{Building Blocks}
            Previous examples can be constructed from a small number of standardized building blocks with different shape changing capabilities $\Delta \alpha = \alpha^+ - \alpha^-$. The required adaptive modules are

            \begin{alignat}{2}\nonumber
                \left\{ \mathbf{A}_{0,4}, \mathbf{A}_{2,6} \right\} &= \left\{ \mathbf{A}_{0^\circ,4^\circ}^{0.1~\textrm{m}}, \mathbf{A}_{2^\circ,6^\circ}^{0.1~\textrm{m}} \right\} \qquad && \Delta \alpha/L = 40^\circ/\textrm{m}\\
                \mathbf{A}_{-22,-16} &= \mathbf{A}_{-22^\circ,-16^\circ}^{0.1~\textrm{m}} \qquad && \Delta \alpha/L = 60^\circ/\textrm{m}\\\nonumber
                \left\{ \mathbf{A}_{-6,2}, \mathbf{A}_{-4,4} \right\} &= \left\{ \mathbf{A}_{-6^\circ,2^\circ}^{0.1~\textrm{m}}, \mathbf{A}_{-4^\circ,4^\circ}^{0.1~\textrm{m}} \right\} \qquad && \Delta \alpha/L = 80^\circ/\textrm{m}.
            \end{alignat}

            \noindent Note that one of the adaptive modules is redundant since $\mathbf{A}_{0,4} \subset \mathbf{A}_{-4,4}$. This redundancy is advantageous since the shape changing capability $\Delta \alpha/L$ of a module is inversely proportional to its stiffness. Hence, it is best to use adaptive modules with the smallest possible shape changing capability. The considered mechanically actuated modules are

            \begin{alignat}{2}\nonumber
                \left\{ \mathbf{M}_{0,80}, \mathbf{M}_{-80,0} \right\} = \left\{ \mathbf{M}_{0^\circ,80^\circ}^{0.1~\textrm{m}}, \mathbf{M}_{-80^\circ,0^\circ}^{0.1~\textrm{m}} \right\} \qquad && \Delta \alpha/L = 800^\circ/\textrm{m}
            \end{alignat}

            \noindent and the rigid modules are

            \begin{align}
                \left\{ \mathbf{R}_0, \mathbf{R}_1, \mathbf{R}_9 \right\} = \left\{ \mathbf{R}_{0^\circ}^{0.05~\textrm{m}}, \mathbf{R}_{\pm 1^\circ}^{0.05~\textrm{m}}, \mathbf{R}_{\pm 9^\circ}^{0.05~\textrm{m}} \right\}.
            \end{align}

            \noindent The connectivity of building blocks for the leading edge are

            \begin{align}
                \mathbf{S} = \left[
                \mathbf{A}_{2,6}^6\ \Box\
                \mathbf{M}^{3}_{0,80}\ \Box\
                \mathbf{A}_{-4,4}^5 \right]
            \end{align}

            \noindent where, for example, $\mathbf{A}_{2,6}^2 = \left[ \mathbf{A}_{2,6} \ \Box\  \mathbf{A}_{2,6} \right]$. The connectivity of building blocks for the trailing edge are

            \begin{align}
                \mathbf{S} =
                \begin{cases}
                    \left[
                    \mathbf{R}_0^6 \ \Box\
                    \mathbf{A}_{0,4}^8 \right] & \mbox{top surface}\\
                    \left[
                    \mathbf{R}_1^{12} \ \Box\
                    \mathbf{A}_{-6,2}^5 \right] & \mbox{bottom surface}
                \end{cases}
            \end{align}

            \noindent and a passenger seat can be constructed by connecting

            \begin{align}
                \mathbf{S} = \left[
                \mathbf{A}_{-6,2}^4 \ \Box\
                \mathbf{A}_{-22,-16}^2 \ \Box\
                \mathbf{R}_9^5 \ \Box\
                \mathbf{M}_{-80,0}\right].
            \end{align}


    \section{Conclusions}
        This article introduced a modular approach to adaptive structures. Adaptive modules consist of compliant pressure actuated cells that are terminated at both ends via rigid connectors. These kind of actuators are lightweight, strong, energy efficient and can be produced from a single material with the help of advanced manufacturing techniques such as injection molding or rapid prototyping. Analytical and numerical models were used to show that cytoskeletons within each cell decrease, increase the weight, stiffness of adaptive modules. Modules possess a fixed arc length and reassemble a circular arc with a varying central angle. An algorithm that breaks down any continuous target shape into a given set of modules was derived. It was shown that it is possible to build an adaptive passenger seat and a leading, trailing edge of an aircraft with only a few standardized modules.



    \newpage
    \appendix

    \section{Coordinates of Leading and Trailing Edge}
    \label{sec:CoordinatesAirfoil}
        Points that describe the shapes of a leading edge for high, low flight speeds are summarized in Table~\ref{tab:LeadingEdge}. Corresponding coordinates for a trailing edge\footnote{Only one set of points is given for the trailing edge since available data is based on a rigid Fowler flap.} are summarized in Table~\ref{tab:TrailingEdge}. Note that this airfoil geometry is similar to the one used by Kintscher and Wiedemann~\cite{Kintscher2012}.
        \begin{table}[htbp]
            \begin{minipage}[c]{0.5\textwidth}
                \begin{center}
                    \scriptsize
                    \begin{tabular}{r|rr|rr}
                        Node  & $x_1$ [mm]  &  $y_1$ [mm] & $x_2$ [mm]  &$y_2$ [mm]\\\hline
                           1  &          0  &     -184.00 &          0  &   -184.00\\
                           2  &      63.35  &     -179.02 &      62.93  &   -181.89\\
                           3  &     107.56  &     -174.33 &     106.64  &	-180.41\\
                           4  &     149.70  &     -169.46 &     148.06  &	-180.17\\
                           5  &     190.38  &     -164.46 &     187.74  &	-181.36\\
                           6  &     228.29  &     -159.25 &     224.42  &	-183.58\\
                           7  &     263.60  &     -153.86 &     258.29  &   -186.65\\
                           8  &     298.19  &     -148.42 &     291.15  &   -191.03\\
                           9  &     330.92  &     -142.87 &     321.95  &   -196.17\\
                          10  &     361.15  &     -137.18 &     350.10  &   -201.50\\
                          11  &     389.36  &     -131.39 &     376.10  &   -206.96\\
                          12  &     415.82  &     -125.51 &     400.24  &   -212.39\\
                          13  &     440.50  &     -119.56 &     422.51  &   -217.54\\
                          14  &     463.47  &     -113.54 &     443.04  &   -222.23\\
                          15  &     484.77  &     -107.45 &     461.89  &	-226.27\\
                          16  &     504.39  &     -101.32 &     479.08  &	-229.45\\
                          17  &     522.36  &      -95.13 &     494.68  &	-231.63\\
                          18  &     538.85  &      -88.90 &     508.86  &	-232.79\\
                          19  &     554.07  &      -82.64 &     521.85  &	-232.98\\
                          20  &     568.20  &      -76.35 &     533.82  &	-232.25\\
                          21  &     581.34  &      -70.04 &     544.85  &	-230.60\\
                          22  &     593.52  &      -63.71 &     555.00  &	-228.02\\
                          23  &     604.61  &      -57.36 &     564.18  &	-224.38\\
                          24  &     614.36  &      -50.99 &     572.21  &	-219.47\\
                          25  &     622.55  &      -44.60 &     578.90  &	-213.12\\
                          26  &     629.16  &      -38.19 &     584.28  &	-205.36\\
                          27  &     634.35  &      -31.77 &     588.50  &	-196.35\\
                          28  &     638.34  &      -25.34 &     591.72  &	-186.30\\
                          29  &     641.30  &      -18.90 &     594.11  &	-175.41\\
                          30  &     643.39  &      -12.46 &     595.79  &	-163.82\\
                          31  &     644.64  &       -6.01 &     596.80  &	-151.63\\
                          32  &     645.03  &           0 &     597.11  &	-139.74\\
                          33  &     644.53  &	     6.87 &     596.71  &	-132.40
                    \end{tabular}
                \end{center}
            \end{minipage}
            \begin{minipage}[c]{0.5\textwidth}
                \begin{center}
                    \scriptsize
                    \begin{tabular}{r|rr|rr}
                              & $x_1$ [mm]  &  $y_1$ [mm] & $x_2$ [mm]  &$y_2$ [mm]\\\hline
                          34  &     643.23  &       13.32 &     595.67  &	-125.05\\
                          35  &     641.25  &       19.76 &     594.08  &	-117.33\\
                          36  &     638.65  &       26.20 &     591.97  &	-109.27\\
                          37  &     635.39  &       32.63 &     589.34  &	-100.86\\
                          38  &     631.42  &       39.06 &     586.12  &	 -92.08\\
                          39  &     626.61  &       45.49 &     582.22  &	 -82.90\\
                          40  &     620.86  &       51.91 &     577.52  &	 -73.25\\
                          41  &     614.03  &       58.32 &     571.93  &	 -63.12\\
                          42  &     606.04  &       64.71 &     565.37  &	 -52.49\\
                          43  &     596.81  &       71.09 &     557.73  &	 -41.37\\
                          44  &     586.25  &       77.45 &     548.95  &	 -29.77\\
                          45  &     574.25  &       83.78 &     538.90  &	 -17.70\\
                          46  &     560.69  &       90.08 &     527.47  &	  -5.19\\
                          47  &     545.63  &       96.34 &     514.66  &	   7.66\\
                          48  &     529.22  &      102.58 &     500.59  &	  20.69\\
                          49  &     511.65  &      108.78 &     485.40  &	  33.76\\
                          50  &     493.00  &      114.95 &     469.12  &	  46.79\\
                          51  &     473.21  &      121.08 &     451.68  &     59.71\\
                          52  &     452.18  &      127.17 &     432.97  &	  72.50\\
                          53  &     429.77  &      133.21 &     412.86  &	  85.11\\
                          54  &     405.88  &      139.20 &     391.20  &	  97.49\\
                          55  &     380.43  &      145.12 &     367.90  &	 109.56\\
                          56  &     353.41  &      150.97 &     342.92  &	 121.23\\
                          57  &     324.91  &      156.75 &     316.31  &	 132.43\\
                          58  &     294.97  &      162.45 &     288.11  &    143.06\\
                          59  &     263.56  &      168.08 &     258.26  &	 153.08\\
                          60  &     230.61  &      173.62 &     226.65  &	 162.46\\
                          61  &     196.10  &      179.06 &     193.30  &    171.15\\
                          62  &     160.06  &      184.40 &     158.19  &	 179.13\\
                          63  &     122.43  &      189.63 &     121.29  &	 186.42\\
                          64  &      83.18  &      194.74 &      82.56  &	 193.01\\
                          65  &      42.43  &      199.74 &      42.16  &	 198.97\\
                          66  &          0  &      204.36 &          0  &    204.36
                    \end{tabular}
                \end{center}
            \end{minipage}
            \caption{Leading edge coordinates of first $\left(x_1,y_1\right)$ and second $\left(x_2,x_2\right)$ target shape.}
            \label{tab:LeadingEdge}
        \end{table}

        \begin{table}[htbp]
            \begin{minipage}[c]{0.24\textwidth}
                \begin{center}
                    \scriptsize
                    \begin{tabular}{r|rr}
                        Node  &   x [mm]  & y [mm]\\\hline
                           1  &        0  &	     0\\
                           2  &    27.11  &	  8.06\\
                           3  &    78.96  &	 13.69\\
                           4  &   110.26  &	 19.35\\
                           5  &   141.37  &	 25.02\\
                           6  &   172.50  &	 30.60\\
                           7  &   203.38  &	 36.37\\
                           8  &   233.95  &  42.08\\
                           9  &   264.58  &	 47.77\\
                          10  &   295.48  &	 53.45\\
                          11  &   326.65  &  59.12\\
                          12  &   358.04  &  64.78\\
                          13  &   389.73  &	 70.41\\
                          14  &   421.98  &	 76.02\\
                          15  &   455.23  &	 81.57\\
                          16  &   490.02  &	 87.02\\
                          17  &   526.84  &	 92.33
                    \end{tabular}
                \end{center}
            \end{minipage}
            \begin{minipage}[c]{0.24\textwidth}
                \begin{center}
                    \scriptsize
                    \begin{tabular}{r|rr}
                        Node  &   x [mm]  & y [mm]\\\hline
                          18  &   566.29  &	 97.46\\
                          19  &   609.27  &	102.29\\
                          20  &   656.78  &	106.67\\
                          21  &   710.06  &	110.32\\
                          22  &   770.05  &	112.69\\
                          23  &   834.26  &	113.18\\
                          24  &   896.83  &	111.64\\
                          25  &   953.57  &	108.56\\
                          26  &  1003.82  &	104.50\\
                          27  &  1048.91  &	 99.87\\
                          28  &  1090.23  &	 94.89\\
                          29  &  1090.23  &	109.43\\
                          30  &  1060.82  & 115.19\\
                          31  &  1031.76  &	120.97\\
                          32  &  1003.11  &	126.76\\
                          33  &   974.87  &	132.58\\
                          34  &   947.01  &	138.42
                    \end{tabular}
                \end{center}
            \end{minipage}
            \begin{minipage}[c]{0.24\textwidth}
                \begin{center}
                    \scriptsize
                    \begin{tabular}{r|rr}
                        Node  &   x [mm]  & y [mm]\\\hline
                          35  &   919.54  &	144.28\\
                          36  &   892.51  &	150.15\\
                          37  &   865.90  &	156.05\\
                          38  &   839.69  & 161.96\\
                          39  &   813.87  &	167.89\\
                          40  &   788.21  &	173.83\\
                          41  &   762.90  &	179.78\\
                          42  &   736.98  &	185.70\\
                          43  &   715.17  &	190.44\\
                          44  &   682.42  &	197.91\\
                          45  &   657.62  &	203.53\\
                          46  &   630.37  &	209.40\\
                          47  &   603.05  &	215.26\\
                          48  &   575.52  &	221.11\\
                          49  &   547.28  &	226.93\\
                          50  &   518.30  &	232.72\\
                          51  &   488.79  &	238.47
                    \end{tabular}
                \end{center}
            \end{minipage}
            \begin{minipage}[c]{0.24\textwidth}
                \begin{center}
                    \scriptsize
                    \begin{tabular}{r|rr}
                        Node  &   x [mm]  & y [mm]\\\hline
                          52  &   458.76  & 244.20\\
                          53  &   428.04  & 249.89\\
                          54  &   396.53  &	255.54\\
                          55  &   364.29  &	261.14\\
                          56  &   331.30  &	266.71\\
                          57  &   297.46  &	272.22\\
                          58  &   262.70  &	277.68\\
                          59  &   226.95  &	283.06\\
                          60  &   190.22  &	288.39\\
                          61  &   152.45  &	293.64\\
                          62  &   113.55  &	298.81\\
                          63  &    73.44  &	303.88\\
                          64  &    32.04  &	308.85\\
                          65  &        0  & 312.48
                    \end{tabular}
                    \vspace{9.6mm}
                \end{center}
            \end{minipage}
            \caption{Trailing edge coordinates of first target shape.}
            \label{tab:TrailingEdge}
        \end{table}
\end{document}